\documentstyle[preprint,aps,epsfig]{revtex}
\newif\iftightenlines\tightenlinesfalse
\tightenlines\tightenlinestrue

\begin{document}
\draft

\title{Tau Neutrinos Underground:
Signals of $\nu_\mu \rightarrow \nu_\tau$ Oscillations with Extragalactic 
Neutrinos}
\author{Sharada Iyer Dutta$^1$, Mary Hall Reno$^{2}$ and Ina Sarcevic$^1$}
\address{
$^1$Department of Physics, University of Arizona, Tucson, Arizona
85721\\
$^2$Department of Physics and Astronomy, University of Iowa, Iowa City,
Iowa 52242
}

\maketitle

\begin{abstract}
\widetext 
The appearance of high energy tau neutrinos due to 
$\nu_\mu \rightarrow \nu_\tau$ oscillations of extragalactic neutrinos 
can be observed by 
measuring the neutrino induced upward hadronic and electromagnetic showers 
and upward muons.  We evaluate quantitatively the tau neutrino regeneration
in the Earth for a variety of extragalactic neutrino fluxes.
Charged-current interactions of the upward tau neutrinos 
below and in the detector, and the subsequent tau decay 
create 
muons or hadronic and electromagnetic showers.  The background 
for these events are muon neutrino and electron neutrino charged-current and 
neutral-current interactions, where in addition to extragalactic neutrinos, we 
consider atmospheric neutrinos.  
We find significant signal to background 
ratios for the hadronic/electromagnetic showers with 
energies above 10 TeV to 100 TeV initiated by the 
extragalactic neutrinos.  
We show that 
the tau neutrinos from point sources also have the potential for
discovery above a 1 TeV threshold.  A
kilometer-size neutrino telescope has a very good chance of detecting 
the appearance of tau neutrinos when both muon and hadronic/electromagnetic
showers are detected.  

\end{abstract}
\vskip 0.1true in


\section{Introduction}

A recent breakthrough in the study of neutrino oscillations came from the 
observation by the 
Super-Kamiokande experiment of a deficit
of upward-going atmospheric muon neutrinos \cite{superk}.  
The observed electron neutrino flux  was found to be 
consistent with the theoretical 
expectation from models of cosmic ray production of neutrinos.
Furthermore, SuperK measurements are 
consistent with earlier experiments \cite{kamioka,kamR,imb,soudan} 
which detected anomalous 
ratios of the $\nu_\mu$ to $\nu_e$ flux.  
The new high-statistics 
data disfavor scenarios 
in which $\nu_\mu$'s oscillate into sterile neutrinos ($\nu_s$) \cite{fgv}, 
and the data are consistent with 
$\nu_\mu$ to $\nu_\tau$ oscillation (99$\%$ CL) with a
large mixing angle, $\sin^2 \theta > 0.84$ and  a 
neutrino mass squared difference
of $2 \times 10^{-3}$ eV$^2 < \Delta m^2 < 6 \times 10^{-3}$ eV$^2$.  

Direct detection of $\nu_\tau$ appearance is extremely 
difficult because  at low energies, the charged-current 
cross section for producing a tau is small and the
tau has a very short lifetime. 
Several long-baseline experiments with accelerator sources
of $\nu_\mu$  \cite{MINOS,K2K,ICARUS,NOE,OPERA} 
have been proposed with the goal of detecting tau neutrinos 
from oscillations, thus confirming the SuperK results.  
The only convincing evidence of neutrino oscillations to date 
involves astrophysical sources, neutrinos from the sun and atmospheric 
neutrinos.  These observations involve indirect measurements, namely 
the disappearance of the expected neutrino fluxes.  

We have recently discussed the possibility of using
a kilometer-size 
neutrino telescope to detect tau neutrinos from
extragalactic sources of high-energy 
neutrinos such as Active Galactic Nuclei (AGN) and Gamma Ray Bursts
(GRB), assuming $\nu_\mu\leftrightarrow \nu_\tau$ with
the oscillation parameters of the SuperK experiment \cite{irs}.  
The 
probability for 
$\nu_\mu \rightarrow \nu_\tau$ is given by \cite{kayser}  
\begin{equation}
P(\nu_\mu \rightarrow \nu_\tau; L) = \rm{sin^2}2\theta \,\,\,\rm{sin{^2}}
\Biggl(\frac{1.27\Delta m^2({\rm eV}^2)L({\rm km})}{E({\rm GeV})}\Biggr)\, .
\end{equation}
Assuming two flavor oscillations,
muon neutrinos produced in AGN or GRB would oscillate to tau neutrinos as 
they travel to the Earth.  
Over astronomical distances
in the range of a megaparsec to thousands of megaparsecs,
by measuring tau neutrino fluxes, one could, in principle, 
probe
oscillations down to
$\Delta m^2 \sim 10^{-17}$ eV$^2$, nine orders of magnitude below
current neutrino experiments \cite{halzen,wspl}.
On the other hand, for the SuperK parameter range, with $\Delta m^2$ on the
order of $10^{-3}$ eV$^2$ and $\sin^22\theta\simeq 1$, the oscillation
probability is 0.5. It is this latter possibility that we explore in this
paper.

We use the simplest assumption for the flavor content of extragalactic
sources of neutrinos, in the absence of oscillations, for the ratio 
$\nu_e:\nu_\mu:\nu_\tau$ to be $1:2:0$. This is based on a counting
argument applied to 
$\pi\rightarrow \mu\nu_\mu$ and $\mu\rightarrow \nu_\mu
e\nu_e$ processes.  
With the two-flavor oscillations suggested by the SuperK experiment,
the flavor ratio becomes $1:1:1$ after the neutrinos have traveled over
astronomical distances. Even in the three-flavor oscillation
scenario, the ratio is still $1:1:1$, because the path length of high
energy extragalactic neutrinos is much larger than any neutrino
oscillation length supported by the solar, atmospheric or accelerator
data \cite{yasuda}.

The ratio for $\nu_e:\nu_\mu$ might get 
modified at high energies due to muon cooling 
\cite{rm}. In addition, $\nu_e$ from neutron decay might give significant 
contribution, resulting in neutrino fluxes  dominated by electron 
neutrinos as in the case of diffuse neutrino fluxes from propagating cosmic 
rays \cite{stanev0003}.  
We comment qualitatively in the discussion
section on how our results are altered with more realistic, flavor-dependent
neutrino energy cutoffs. Regardless of the flavor content of the source,
the maximal mixing suggested by the SuperK experimental results mean that there
will be an appreciable tau neutrino component at the Earth, so one
is interested in tau neutrino detection in high energy
neutrino telescopes such as ANTARES \cite{antares}, NESTOR \cite{nestor},
AMANDA \cite{amanda} and the next generation of large
underground detectors \cite{km3}.

Tau neutrino detection requires an understanding of the effect
of propagating through the Earth on the tau neutrino flux.
The propagation of ultra-high energy 
tau neutrinos through the Earth is quite different from 
muon and electron neutrinos. The Earth 
never becomes opaque to tau neutrinos, while 
muon and electron neutrinos 
are absorbed via charged-current interactions before 
reaching the opposite
surface \cite{halzen}.
Ultrahigh-energy tau neutrinos  
interact in the Earth producing taus
which, due to the short lifetime, decay back into
tau neutrinos with lower energy.  This cascade continues until the 
tau neutrinos reach the detector on the opposite side of the Earth
or until the energy of the neutrinos is small enough that the
interaction length of the neutrino is longer than the path length
through the Earth. 
The energy and nadir angle dependence of the extragalactic tau
neutrinos fluxes
have been examined quantitatively in
Refs. \cite{irs,bottai}. For certain fluxes, those that do 
not decrease too steeply with energy, there are significant enhancements
of the tau neutrino flux relative to the muon neutrino flux at energies
below $\sim 10^6$ GeV.

An enhancement of the tau neutrino flux does not necessarily translate
to dramatic modifications of the standard model
(no-oscillation) rates for upward-going muons, especially in view of
uncertainties in the normalization of the extragalactic fluxes.
However, by comparing rates for upward-going muons with rates for
upward hadronic/electromagnetic (EM) showers, the signature of
tau neutrino interactions is unambiguous for a large range of
neutrino fluxes. In the next section, we briefly introduce
the extragalactic  and generic  $\nu_\tau$ fluxes
$F_{\nu_\tau}^{o}\sim E^{-n}$ for $n=1,2$
that are used here. After reviewing neutrino propagation through
the Earth, we describe
$\nu_\tau$ signatures. 
The fluxes considered here have a range of energy behaviors.
Even if the normalizations of the neutrino fluxes 
are uncertain, and in some cases optimistic,
it is useful to make quantitative comparisons of the event rates
for upward muons and upward hadronic/EM showers,
with and without neutrino oscillations, 
which we do in Section IV. 
The quantitative results for specific models 
lead to model independent conclusions, which we summarize
graphically. 
Tau neutrino appearance would provide an independent confirmation of 
the SuperK results and would point towards the better understanding of 
physics beyond the Standard Model.

\section{High Energy Neutrinos Sources}

Active Galactic Nuclei are the most luminous objects in the Universe.  
Most of this radiation comes from their central region, indicating that 
the energy radiation most likely comes 
from accretion 
of matter into a superheavy black hole.  Protons within the AGN 
may get accelerated via 
first order Fermi acceleration to very high energies.  They interact with 
protons and photons in the infalling gas, or they may exist
in the jets along the 
rotation axis and interact with photons there.  
Photon-proton and proton-proton interactions produce pions, which 
decay into 
charged leptons, neutrinos and photons.  Energetic photons 
($E_\gamma \sim 100$ MeV)
from about 40 AGN observed by the EGRET collaboration \cite{EGRET}, and 
 TeV photons have been detected from
Mkn 421, Mkn 501 \cite{AGN} and 1ES2344+514 
 \cite{Weekes}.  Although these photons are conventionally explained by 
inverse Compton 
scattering from energetic electrons, this explanation is not 
without problems, and a hadronic origin of gamma-ray photons from AGN is a 
viable alternative [27].  If a large fraction of the observed energy in 
high energy photons from AGN is produced in hadronic interactions, then 
AGN are also 
powerful sources of
ultrahigh-energy (UHE) neutrinos \cite{stecker,Mannheim95,gqrs96,gqrs98}.

In Fig. 1 we show neutrino fluxes predicted in the AGN models of
Stecker and Salamon \cite{stecker} and Mannheim Model A \cite{Mannheim95}.
Both of these models predict neutrinos fluxes that 
represent the upper bounds for their class of the 
models.  
In particular, the Stecker-Salamon flux is an upper bound for AGN core 
emission, while Mannheim Model A 
is an upper bound for 
AGN jet emission models.  Stecker-Salamon flux is bound by the 
the diffuse X-ray background, while Mannheim flux is bound by the 
extragalactic gamma ray background.  
The steep, low energy neutrino flux in Mannheim's model is the emission 
from 
the host galaxy via 
$pp$ interactions of the AGN protons in the galactic gas disk.  
Since this part of the flux is derived with the assumption that 
all protons end up in the disk, it should be regarded as an upper 
bound.  
Stecker-Salamon flux at energies above 1 PeV may get reduced due to the 
cooling of pions and muons in the strong magnetic fields of 
AGN cores \cite{rm}.  
The fluxes plotted in Fig. 1 are for the sum of muon 
neutrino plus antineutrino,
at the source, namely, without accounting for oscillation over
astronomical distances. We label the fluxes in the absence of
oscillations by $F^s_{\nu+\bar{\nu}}$.

Another extragalactic source with powerful radiation and possibly 
associated high energy neutrino flux are the 
gamma ray bursts (GRB).  
Several models have been proposed in order to explain the origin of GRB's 
\cite{othergrb,fireball}.  
In the fireball model \cite{fireball}, the gamma ray bursts 
are produced by the dissipation of the kinetic energy of the 
relativistic expanding fireball with 
a large fraction,  $>10\%$,  of the fireball energy
being converted by photopion production to high energy neutrinos \cite{waxman}.
Photomeson production takes place when extremely energetic protons 
accelerated  
at high energies in the ultra-relativistic shocks interact with synchrotron 
photons inside the fireball. The decay of these charged pions 
and subsequently produced muons then 
produce electron and muon neutrinos.  Contributions from 
proton-proton collision can be neglected in this model.  
In Fig. 1 we show the neutrino fluxes for gamma ray burst model 
of Waxman and Bahcall (GRB\_WB)\cite{bounds}, in which they parameterize
the flux by 
\begin{eqnarray}
F^s_{\nu+\bar{\nu}}(E) = 4.0 \times 10^{-\alpha} E^{-n}, \nonumber
\end{eqnarray}
\noindent
where $\alpha=13$ and $n=1$ for $E < 10^5$ GeV and 
$\alpha=8$, $n=2$ for $ 10^5 < E < 10^7$ GeV and $\alpha=1$, $n=3$ for 
$E>10^7$ GeV.

Theoretical work has been done to set upper bounds on high energy
neutrino fluxes from AGN jets and GRB \cite{bounds}. 
The bounds are based on the
theoretical correlations between the cosmic
ray flux and/or the extragalactic gamma ray flux and the neutrino flux. 
These bounds have some model dependence, and they tend to be weaker
in the range of energies considered here than at higher energies
($E\sim 10^7-10^9$ GeV). The AGN and GRB neutrino fluxes used here 
satisfy 
these bounds.  

Cosmic topological defects (TD) such as magnetic monopoles, cosmic strings and 
domain walls are predicted to be formed in the Early Universe as a result 
of symmetry breaking and phase transition in Grand Unified Theories (GUTs) 
of 
particle interactions.
In the TD models, $\gamma$-rays, electrons
(positrons), and neutrinos are produced directly at ultra-high energies 
via 
cascades initiated by the decay of a supermassive elementary ``X''
particle associated with some Grand Unified Theory, rather 
than being produced in high
energy hadronic interactions.
The X particle is usually thought to be released from topological
monopoles left over from GUT phase transition. It 
decays into quarks, gluons, leptons.
In this paper, we consider neutrino fluxes from topological defects models of 
Sigl-Lee-Schramm-Coppi (TD\_SLSC) \cite{Sigl} and the model of 
Wichoski-MacGibbon-Brandenberger (TD\_WMB) \cite{wmb}.  The main difference 
between these two models is the main channel for energy loss of the string 
network, in the former it is the gravitational radiation, while in the later 
it is the particle production. 
Both of these fluxes should be regarded as upper limits for TD models, 
because they have been constructed in such a way to satisfy the 
bound imposed by the measured cosmic ray and gamma ray fluxes \cite{ps96}.  
These fluxes are shown
in Fig. 1, where we take representative flux of WMB model 
with the string mass parameter giving the largest neutrino flux that is 
consistent with cosmic ray data.  This flux is also below the 
Frejus \cite{frejus} and Fly's Eye \cite{fly} experimental limits on
the neutrino flux.  

We also consider two generic fluxes that have a power law behavior.
The flux
\begin{equation}
F^{s}_{\nu+\bar{\nu}}(E)=10^{-7}{(E/{\rm GeV})^{-2}} 
(\rm{cm}^{-2} \rm{s}^{-1}\rm{sr}^{-1}\rm{GeV}^{-1})\ 
\end{equation}
gives numerically stable results, however, our
calculations with a flux with $F^s\sim E^{-1}$
is unstable at very high energies. Consequently, we use
\begin{equation}
F^{s}_{\nu+\bar{\nu}}(E)=10^{-13}{(E/{\rm GeV})^{-1}}{1\over
(1+E/10^8 {\rm GeV})^2 }\ 
(\rm{cm}^{-2}\rm{s}^{-1}\rm{sr}^{-1}\rm{GeV}^{-1})
\end{equation}
as a way to cutoff the high energy behavior.  We show
results for attenuated fluxes for neutrino energies up to $10^6$ GeV.
We have chosen the multiplicative factors in $E^{-1}$ and $E^{-2}$  fluxes in 
such a way that they exceed atmospheric flux at neutrino energies between 
10 TeV and 100 TeV.  
The upper bound for strong source evolution discussed 
recently by Waxman and Bahcall \cite{bounds} would correspond to a limit of 
$2\times 10^{-8} E^{-2}$ (in the same units as Eq. (2)), 
a factor of 5 smaller than the 
choice of normalization 
we take in this paper.  

We also show in Fig. 1 the atmospheric neutrino flux at zenith angle of $0^{\circ}$ and 
the horizontal flux.  
In our evaluation of the atmospheric backgrounds, we use 
the  atmospheric muon and electron neutrino fluxes as a function of 
zenith angle \cite{agls}.  

We do not consider the 
neutrino flux 
from cosmic ray interactions with the microwave background. This diffuse
neutrino flux is typically present at energies higher than we
consider here \cite{cmb}, and it gives 
low event rates \cite{gqrs96}.
Furthermore, the cosmic ray interactions with the solar atmosphere 
are another source
of neutrinos, however, for energies above a TeV, the flux scales
as $E^{-3}$ or steeper \cite{sun}. As we see below, tau neutrino
regeneration will not be a very important feature in fluxes with
such large spectral indices.

\section{Tau Neutrino Propagation through the Earth}

For neutrino energies above 1 TeV, the oscillation probability for
$\nu_\mu\rightarrow\nu_\tau$ in the Earth
is less than a percent for the parameters
constrained by the SuperK experiment. As a consequence, we can neglect
neutrino oscillation in our evaluation of tau neutrino propagation
accounting for interactions in the Earth.

The coupled transport equations for the fluxes of 
the tau neutrino and its charged partner are given by 
\begin{eqnarray}
& &\frac{\partial F_{\nu_{\tau}}(E,X)}{\partial X} =-
\frac{F_{\nu_{\tau}}(E,X)}
        {\lambda_{\nu_{\tau}}(E)}
+ \int_E^\infty dE_y
G_{\nu_\tau\rightarrow \nu_\tau}(E_y,E,X)\\ \nonumber
&& + \int_E^\infty dE_y
G_{\tau\rightarrow \nu_\tau}(E_y,E,X)
\end{eqnarray}
and for taus as,
\begin{eqnarray}
 & & \frac{\partial F_\tau(E,X)}{\partial X} = 
-\frac{F_\tau(E,X)}{\lambda_\tau(E)}
        - \frac{F_\tau(E,X)}{\lambda_\tau^{dec}(E,X,\theta)}\\ \nonumber
&& + \int_E^\infty dE_y
G_{\tau\rightarrow \tau}(E_y,E,X)
+\int_E^\infty dE_y
G_{\nu_\tau\rightarrow \tau}(E_y,E,X).
\end{eqnarray}

Here 
$F_{\nu_{\tau}}(E,X)$ and $ F_\tau(E,X)$ are differential energy
spectrum 
of tau neutrinos and tau
respectively, for lepton energy $E$, 
at a column depth $X$ in the medium defined by
\begin{equation}
X = \int_0^L\rho(L')dL'.
\end{equation}
The density of the medium a distance $L$ from the Earth-atmosphere
boundary, measured along the neutrino beam path, is
$\rho(L)$.
The lepton interaction length (in g/cm$^2$) is
$\lambda(E)$ and
$\lambda_{\tau}^{dec}(E,X,\theta)$ 
is the decay length of the tau.  

The functions $G_{i\rightarrow j}$ schematically represent interaction
or decay contributions to lepton $j$ from lepton $i$.
We limit our evaluations of the tau neutrino flux to $E_{\nu_\tau}<10^6$ GeV.
Consequently, we can ignore several terms in the coupled differential
equations: the term with $G_{\tau\rightarrow \tau}$ and
the term $-F_\tau/\lambda_\tau$, both in Eq. (5). Only the
decay contribution to the last term in Eq. (2) 
($G_{\tau\rightarrow \nu_\tau}$) is included in our evaluation. This is
justified by the fact that
the tau decay is significantly more important
than interactions for the energy range of interest, namely $E<10^6$ GeV.
The neglected
terms start contributing for lepton energies on the order of $10^8$ GeV.
Detailed formulae  for $G_{i\rightarrow j}$ appear in Ref. \cite{irs}.

In our previous work, we have described an analytic method for 
solving these transport equations \cite{irs},
based on the method of Naumov and Perrone \cite{Nauper}.  
We have evaluated 
the upward $\nu_\tau$ flux for a selection of initial fluxes \cite{irs}.  
We have shown that 
for ``flat'' initial neutrino fluxes
($F\sim E^{-1}$), a significant number of
high energy $\nu_\tau$'s cascade down in 
energy, resulting in enhanced low energy flux relative to the attenuated 
$\nu_\mu$ flux.  Here, we evaluate the tau neutrino flux for a
more comprehensive selection of incident fluxes, including both
neutrino and antineutrino attenuation. In all of our results below, we
evaluate the sum of neutrino plus antineutrino fluxes or rates.

In Fig. 2 and 3, 
we show the attenuated tau neutrino plus antineutrino 
flux (blue line) 
and attenuated muon
neutrino plus antineutrino flux (Red line), scaled by a factor of
the neutrino energy $E$,
assuming the equal fluxes
of tau neutrinos and muon neutrinos incident on the surface of the Earth
at a nadir angle of $0^{\circ}$ for 
$F^{o}_{\nu+\bar{\nu}}(E_\nu)=0.5\times 10^{-13}{E^{-1}}$, 
$F^{o}_{\nu+\bar{\nu}}(E_\nu)=0.5\times 
10^{-7}{E^{-2}}$, the Stecker-Salamon AGN model, the
Mannheim AGN (Model A), 
the two topological defects models, 
the 
Waxman-Bahcall GRB model and 
the atmospheric flux.  We note that the enhancement of the tau neutrino 
flux relative to the initial flux and also to the muon neutrino flux, 
is 
prominent for the flat fluxes, such as 
$F^{o}_\nu(E_\nu)\sim E^{-1}$, the Stecker-Salamon AGN model and 
the
topological defects model of Sigl {\it et al}.  In case of the atmospheric 
flux, which represents the background, the enhancement is very small 
due to the steepness of the initial neutrino flux.  

The angular dependence of the upward $\nu_\tau$ flux is also 
distinct. As an example,
in Fig. 4, for the AGN model of Mannheim (Model A) \cite{Mannheim95}, we show
the ratio of the neutrino flux scaled by the flux at $X=0$ for two
nadir angles, $\theta=0^{\circ}$ and $\theta=30^{\circ}$, as a function
of neutrino energy. Because of the shape of the 
initial flux, steep 
for energies below $10^6$ GeV, and flat for higher energies, 
the enhancement of the 
tau neutrino flux becomes significant only for energies above $10^6$ GeV.  
At fixed energies of $10^4$, $10^5$, and $10^6$ GeV,
we show the same flux ratios as a function of nadir angle.

\section{Detection of $\nu_\tau$ appearance}

Detection of muon neutrinos, in general,
is via their charged-current interactions.  Produced muons have
very large average range
making the effective volume of an underground detector
significantly larger than the instrumented volume.  On the other hand, tau
neutrino charged-current interactions produce tau, which has a very
short lifetime,
making its detection extremelly difficult.  Only at
very high energies, $E_\nu > $  PeV, the
production and decay vertices are
separated by a measurable distance providing a
distinctive signature of tau neutrinos (``double-bang'' events) 
\cite{learned}.   
However, the predicted neutrino fluxes are low at these energies.
For $\nu_\tau$'s
in the energy range of $10^3-10^6$ GeV considered here, 
the produced tau decays after a very short
pathlength back to $\nu_\tau$ plus leptons or hadrons.  Tau neutrinos
will interact via neutral currents, producing a hadronic signal
as well.
Therefore, the signals of tau neutrino interactions
below the double-bang threshold are
muons from tau decay, or hadronic/EM  showers from the tau
production and/or decay.  In the first case the 
background to tau production of high energy muons 
is $\nu_\mu$ charged-current interactions.  In the latter case, the 
backgrounds are $\nu_\mu$ neutral current
and $\nu_e$ charged-current and neutral current 
interactions.  The background rates shown below are with the assumption that
the electromagnetic shower from $\nu_e\rightarrow e$ charged current
interactions cannot be distinguished from hadronic showers. As a consequence,
we evaluate the hadronic/electromagnetic (EM) shower rates.

We assume in the analysis presented below 
that the $\nu_\mu$ charged-current events 
and $\nu_\tau\rightarrow \tau\rightarrow \mu$ events can be rejected
from the contained hadronic/electromagnetic shower signal.
In both cases there is a hadronic shower which includes muons, however,
the muons in the hadronic showers from pion and kaon decays are
significantly less energetic than the muons from $\nu_\mu\rightarrow
\mu$ and $\nu_\tau\rightarrow\tau\rightarrow \mu$. In the latter case,
the energy of the shower is $\sim 1/2$ the incident neutrino energy $E_\nu$
and the energy of the muon is $\sim 1/6-1/2\, E_\nu$.  
On the other hand, 
muons coming from particle decays in the hadronic shower are 
considerably less energetic because of large particle multiplicities.
The average charged particle multiplicities for hadronic interactions at
$\sqrt{s}>40$ GeV are larger than $\sim 10$ particles \cite{isr}, 
so individual muon energies
from charged pion and kaon decays are 
less than $\sim 5\%$ of the
incident neutrino energy. The hadronic shower and very energetic
muon of the ``muon signal'' should stand out 
in comparison to the hadronic/EM shower signal in a detector with good
energy resolution like the proposed kilometer-cubed detector IceCube\cite{ice}.

We describe the evaluation of the muon and hadronic/EM shower event
rates. The event rates for $\nu_\tau\rightarrow \tau\rightarrow \mu$
and $\nu_\mu\rightarrow \mu$ are evaluated and compared with the
no oscillation rates. We evaluate the hadronic/EM shower rates for
signal and background, then compare with the hadronic/EM shower rates
assuming no oscillations of $\nu_\mu$. The relative rates of
muons and hadronic/EM showers prove to be the most effective diagnostic
to neutrino oscillations with the SuperK parameters.

\subsection{Muon Event Rates}

The standard evaluation of the muon event rate per solid angle for
neutrino interactions with
isoscalar nucleons $N$
($\nu_\mu N\rightarrow \mu X$) follows from the
formula \cite{gqrs96}
\begin{eqnarray}
{\rm Rate} & &  = A N_A  \int_{E_\mu^{\rm min}}^\infty 
dE_{\nu}\int
dy
\langle R_\mu(E_\nu(1-y), E_\mu^{\rm min})\rangle
{d\sigma_{cc}(E_{\nu},y)\over {dy}}\\ \nonumber
 & & \times  F_{\nu} (E_\nu, X)
\Theta (E_\nu(1-y)-E_\mu^{\rm min}).
\end{eqnarray}
where $y$ is the neutrino energy loss,  
$y=({E_\nu-E_\mu})/{E_\nu}$, and  
${d\sigma_{cc}(E_{\nu},y)}/{dy}$ is the charged current differential 
cross section.  
$F_{\nu}(E_\nu,X)$ is the upward neutrino or antineutrino
flux which
depends on angle implicitly through the pathlength $X$. We assume that
the initial fluxes of muon neutrinos and antineutrinos that reach the
Earth are equal, their sum in the oscillation scenario being half of the
muon neutrino plus antineutrino flux produced at the source.
The fluxes of neutrinos and antineutrinos at the detector are different
because of the difference in charged and neutral current cross sections
below energies of $10^6$ GeV \cite{gqrs96,gqrs98}, however, for
these energies and fluxes, the antineutrino event rates differ
for the neutrino event rates by at most $\pm 20\%$.
The average range of a muon, 
$\langle R_\mu(E_\mu, E_\mu^{\rm min})\rangle$, is the range of a muon produced in a 
charged-current interaction with energy $E_\mu$ 
which, as it passes through the medium, 
looses its energy via bremsstrahlung, ionization, 
pair production and photonuclear interaction 
and arrives in a detector with an 
energy
above $E_\mu^{\rm min}$.  
Avogadro's number is $N_A$ and $A$ is the effective
area of the detector. All of the
event rates calculated are for the sum of neutrino plus antineutrino
contributions to $\mu^+ + \mu^-$ production.

The rate for muons produced by the tau neutrino charged current interactions 
followed by the tau leptonic decays is given by a modified equation,
taking into account the branching fraction for $\tau\rightarrow \nu_\tau
\nu_\mu \mu$ and the decay distribution of the muon
via ${{dn(E_{\tau})}/{dz}}$, where $z=E_\mu/E_\tau$.
The decay formulae used here are listed in the Appendix A.  
The differential event rate is
\begin{eqnarray}
{\rm Rate } & &
= A N_A  \int_{{E_\mu^{\rm min}}}^\infty dE_{\nu}\int dy\int dz
\langle R_\mu(E_\nu(1-y)z, E_\mu^{\rm min})\rangle {\frac{dn(E_\nu(1-y)z)}{dz}}
\\ \nonumber
& & \times
{\frac{d\sigma_{cc}(E_{\nu},y)}{dy}} F_{\nu}
(E_\nu, X)\Theta (E_\nu(1-y)z-E_\mu^{\rm min}).
\end{eqnarray}

In Fig. 5 we show the 
neutrino processes that contribute to the muon production.  
In our evaluation of the
muon event rates we use the
Earth
densities of the
Preliminary Earth Model (PREM) described in Ref. \cite{profile}.
We have used the PREM to determine an average density for a given
nadir angle, then used that average density to evaluate the attenuated
fluxes. We use the muon range evaluated by Lipari and Stanev \cite{lipari}.
The
neutrino and antineutrino cross sections
have been evaluated using the CTEQ5 parton distribution functions \cite{cteq5}.
The effective area $A$ is taken to be 1 km$^2$.

In Figs. 6-9, we show muon event rates for $F^{o}_{\nu+\bar{\nu}}\sim
E^{-1}$, 
$F^{o}_{\nu+\bar{\nu}}\sim E^{-2}$, 
AGN\_SS, AGN\_M95, TD\_WMB, TD\_SLSC and  GRB\_WB 
for $E_\mu^{\rm min}=1,\ 10,\ 100$ TeV.  Blue lines correspond to the 
upward $\mu^+ +\mu^-$ events from $\nu_\tau+ 
\bar{\nu}_\tau+\nu_\mu+\bar{\nu}_\mu$
charged-current interactions (including $\tau \rightarrow \mu$ decay), while 
the red lines are the background contribution from
$\nu_\mu+\bar{\nu}_\mu$  charged-current interaction only.  
We note that the muon enhancement due to the tau neutrino contribution for 
$E^{-1}$ flux is almost factor of 2 for small angles and 25\%
for large angles with
$E_\mu^{\rm min}=1$ TeV. The enhancement is less pronounced at small
nadir angles for increasing threshold energies, for example, the
blue line is about 60\% enhanced relative to the red line
for $E_\mu^{\rm min}=10$ TeV for the $E^{-1}$ flux in Fig. 6. A 
similar enhancement is present for the TD\_SLSC.
For steeper fluxes, such as AGN\_SS, AGN\_M95 and $E^{-2}$, 
the enhancement due to tau neutrino contribution is much smaller, of 
the order of 20-25\%.
  
The muon event rates from the atmospheric neutrino background are
shown in Fig. 10a). The input flux is the angle dependent muon neutrino flux
of Agrawal {\it et al.} \cite{agls}.  
The atmospheric tau neutrino flux is very low, as the 
tau neutrinos are
produced in the atmosphere by comic ray interactions 
with nuclei in the atmosphere, which produce 
$D_s$ whose leptonic decay,  
$D_s\rightarrow \tau \nu_\tau$, gives $\nu_\tau$ \cite{pasq}. 
The rates for the atmospheric tau neutrinos are shown in Fig. 10b).
In the evaluation of the event rates,
we neglect oscillations of atmospheric neutrinos as they travel to 
the Earth and the oscillations through the Earth 
since the oscillation probabilities are small
above our minimum energy of 1 TeV.

The atmospheric neutrino  events represent a background for detection
of extragalactic neutrinos. 
For a muon energy threshold of 1 TeV, the background 
is large, $400-2000$ events per year per steradian for 1 km$^2$ effective
area detector. For a muon threshold of  $E_\mu^{\rm min}=10$ TeV, the
event rates range between $6-80$ events per year per steradian.
A comparison of the event 
rates from Figs. 6-9 with the atmospheric muon neutrino 
background indicates that 
detection of neutrinos from AGN might be possible with 
$E_\mu^{\rm min}=10$ TeV or $100$ TeV.  

The rates for muon events shown in Figs. 6-9 come from assuming
that the tau neutrino and muon neutrino fluxes are equal and
are half the flux of muon neutrinos produced at the source.
Testing the oscillation hypothesis with muon neutrinos alone will be
difficult. We see that with the exception of the $E^{-1}$ and 
TD\_SLSC fluxes, the observed muon rate is about half of what
one would expect in the absence of oscillations. Given the uncertainties
in the normalizations of the predicted fluxes, this factor would
not unambiguously signal the presence of tau neutrinos from oscillations.
The situation with the $E^{-1}$ and 
TD\_SLSC fluxes is only slightly better. There, in the oscillation scenario,
the measured muon event rate is about 80\%  of the no oscillation prediction
at $\theta=0^{\circ}$, 
but less than 70\%\ of the prediction for horizontal events.
Testing the oscillation hypothesis 
by measuring upward 
muons only will be very difficult.

The relatively small contribution to the muon rate
from $\nu_\tau$'s, despite the fact that the attenuated flux of tau
neutrinos is larger than that of the muon neutrinos, is due to the
fact that the muon carries a small fraction of the initial tau neutrino
energy. Consequently, for a muon of a given energy, if it comes from
a tau neutrino (which interacted producing a tau that subsequently decayed
to a muon), the initial tau neutrino has a much higher energy than
a muon neutrino which produces a muon directly via
the charged current $\nu_\mu N\rightarrow \mu X$
process. All predicted neutrino fluxes decrease with energy.
Even with some ``pileup,'' the tau neutrino fluxes are decreasing fast
enough that the muon energy fraction results in sampling a much smaller
tau neutrino flux than the corresponding muon neutrino flux.
It is this observation that leads one to consider signals that
carry a much larger fraction of the incident tau neutrino energy.

\subsection{Upward Hadronic/Electromagnetic Showers and Their Detection}

The hadronic/EM shower signal of $\nu_\tau$ interactions is a much more
promising final state from the theoretical point of view than the
muon signal. The benefit is that the hadronic showers include both
production hadrons and tau decay hadrons, so there is a much higher
fraction of the incident tau neutrino energy visible in the detector 
\cite{stanev}.
The next generation of neutrino telescopes may not be able to 
distinguish between  hadronic and electromagnetic showers, so
we include in the signal and in the 
background, processes that include hadrons
and electron. As mentioned above, we assume that the high energy
muon associated with the target jet in $\nu_\mu$ charged current interactions
will be used to veto the  process $\nu_\mu N\rightarrow \mu X$.
Distinguishing electromagnetic from hadronic showers 
might be possible by looking at the 
difference 
between the front to back ratio of the cascade  
Cherenkov light, and perhaps by the number of residual $\pi\rightarrow \mu 
\rightarrow e$ decay, although this is considered to be 
experimentally difficult 
\cite{john}.  

The processes that go into our evaluation of $\nu_\tau\rightarrow$ hadrons
are
\begin{eqnarray}
\nonumber
& & \nu_\tau N\rightarrow \tau + {\rm hadrons},\ \tau \rightarrow \nu_\tau 
+ {\rm hadrons}\ ,
\\
\nonumber
& & \nu_\tau N\rightarrow \tau + {\rm hadrons}, \tau
\rightarrow \nu_\tau +e +\nu_e\ ,
\\
\nonumber
& & \nu_\tau N\rightarrow \nu_\tau + {\rm hadrons}\ .
\end{eqnarray}
For the charged-current interactions, the 
hadronic/electromagnetic energy is the sum of the energy carried
by the hadrons in tau production, as well as the tau decay hadronic
energy or tau decay electron energy. 

The background 
for the hadronic/electromagnetic showers is due to the $\nu_\mu$ and $\nu_e$ 
neutral current interactions, and $\nu_e$ charged-current interactions are
\begin{eqnarray}
\nonumber
& &  {\nu_{\mu,e}} + N \rightarrow {\nu_{\mu,e}}  + {\rm hadrons} \ ,
\\ \nonumber	
& &  {\nu_e} + N \rightarrow e + {\rm hadrons} \ .
\end{eqnarray}	
For the $\nu_e$ flux, we assume it is equal to the $\nu_\mu$ flux
in the SuperK oscillation scenario.
All of the processes that contribute to the hadronic/EM showers 
are shown in Fig. 11.

The tau neutrino shower event rate per unit solid angle from charged-current
interactions followed by the tau hadronic decay is given by 
\begin{eqnarray}
{\rm Rate} & &  = 
V N_A { \int_{\small{E_{\rm shr}^{\rm min}}}^\infty dE_{\nu}\int dy\int dz
{\frac{dn(E_{\tau})}{dz}}{\frac{d\sigma_{cc}(E_{\nu_\tau},y)}{dy}} 
F_{\nu_\tau}
(E_{\nu_{\tau}}, X)} \\ \nonumber
& & \times \Theta (E_{\nu_\tau}(y + (1-y)(1 - z)) - E_{\rm shr}^{\rm min}).
\end{eqnarray}
The hadronic energy from the broken nucleon 
$ E_{\rm shr}^{int} = E_{\nu}y$ and the hadronic energy from the
decay $ E_{\rm shr}^{\rm decay} = E_{\nu}(1-y)(1 - z)$ are added to get
the total shower energy. Again, $y=(E_\nu-E_\tau)/E_\nu$ for incident
neutrino energy $E_\nu$, while $z=E_\nu^\prime/E_\tau$, where $E_\nu^\prime$
is the energy of the neutrino from the tau decay.  
The differential distributions for the hadronic decay modes are shown
in the Appendix A.  
For the electronic decay of the tau, the differential distribution $dn/dz$
is replaced by the purely leptonic distribution in terms of $z^\prime\equiv
E_e/E_\tau$. The theta function is replaced by
\begin{equation}
\Theta(E_{\nu_\tau}(y+(1-y)(1-z))-E_{\rm shr}^{\rm min})
\rightarrow 
\Theta(E_{\nu_\tau}(y+(1-y)z^\prime)-E_{\rm shr}^{\rm min}) \ .
\end{equation}

The neutral current background event rate is given by
\begin{equation}
{\rm Rate}   = 
V N_A { \int_{\small{E_{\rm shr}^{\rm min}}}^\infty
dE_{\nu}\int dy{\frac{d\sigma_{nc}(E_{\nu},y)}{dy}} F_{\nu}
(E_\nu,X)}\Theta (E_{\nu}y-E_{\rm shr}^{\rm min})\ ,
\end{equation}
while the electron neutrino charged current background rate is given by
\begin{equation}
{\rm Rate}   = 
V N_A { \int_{{E_{\rm shr}^{\rm min}}}^\infty
dE_{\nu}\int dy{\frac{d\sigma_{cc}(E_{\nu},y)}{dy}} F_{\nu}
(E_\nu,X)}\Theta (E_{\nu}-E_{\rm shr}^{\rm min}).\
\end{equation}

In Figs. 12-15 we show the upward hadronic/EM shower
event rates as a function of the nadir angle for 
$E_{\rm shr} > E_{\rm shr}^{\rm min}$ where 
$E_{\rm shr}^{\rm min} =  1$ TeV, 10 TeV and 100 TeV 
for input fluxes:
$F_{\nu+\bar{\nu}}^0 \sim E^{-1}$,
$F_{\nu+\bar{\nu}}^0 \sim E^{-2}$ AGN\_SS, AGN\_M95, TD\_WMB,
TD\_SLSC and GRB\_WB, all assuming that $V=1$ km$^3$.  
The blue lines correspond to the event rates 
from $\nu_\tau+\bar{\nu}_\tau+\nu_e+\bar{\nu}_e$ 
charged-current interactions (and $\tau\rightarrow \nu_\tau+$ hadrons
decay) and from $\nu_\tau+\bar{\nu}_\tau+\nu_\mu+\bar{\nu}_\mu+\nu_e+ 
\bar{\nu}_e$ neutral current interactions.
The red lines are the contributions from 
$\nu_\mu+\bar{\nu}_\mu$ neutral current interaction and 
$\nu_e+\bar{\nu}_e$ charged and neutral current interactions. 
We do not include $\nu_\mu+\bar{\nu}_\mu$ charged-current
interactions in our calculation because these events can be vetoed
by the high energy muons produced in the interactions.
All of the rates shown in these figures assume equal neutrino and
antineutrino fluxes. They are performed in the oscillation scenario
where the ratios of the fluxes $\nu_e:\nu_\mu:\nu_\tau$ are
$1:1:1$.

From Fig. 12a) we note that in the case of the $E^{-1}$ flux, 
the contributions from tau neutrinos are large, a factor of 
4 times larger than the muon neutrino plus electron neutrino 
contribution at zero nadir angle.  
For horizontal showers, the enhancement factor is smaller, about 2 for 
all the energy thresholds that we consider.  
Similarly, for $E^{-2}$ flux, the tau neutrino contribution is 
a factor of 1.7 times larger than the muon neutrino plus electron neutrino 
contributions for upward 
showers.  

Similar conclusions can be drawn from the plots of the other fluxes.
The shower event rates including $\nu_\tau+\bar{\nu}_\tau + \nu_\mu
+\bar{\nu}_\mu+\nu_e+\bar{\nu}_e$ are significantly enhanced
relative to the rates from $\nu_\mu+\bar{\nu}_\mu+\nu_e+\bar{\nu}_e$
in the oscillation scenario. The AGN\_SS rates at zero nadir angle
are comprised of 60\%  tau neutrino induced, decreasing to about 40\% 
tau neutrino induced for 
horizontal showers, as shown in Fig. 13a).  
AGN\_SS flux gives 25-80 shower events for $E_{\rm shr}^{\rm min}=10$ TeV and 6-45 events for
$E_{\rm shr}^{\rm min}=100$ TeV with negligible atmospheric background.
In Fig. 13b) we show event rates for 
AGN\_M95 model.   We find 3-6 shower events per year per steradian for
$E_{\rm shr}^{\rm min}=10$ TeV, 
with atmospheric background of 2-16 events.
Detection of
events with higher energy threshold would require looking at almost horizontal
showers, where the background is small.

The TD\_WMB model in Fig. 14a) shows an enhancement of between
2.1-2.3 for zero nadir angle, and a factor of 1.7 
for almost horizontal showers. Fig. 14b) shows the more striking enhancement
in the TD\_SLSC model, where the enhancement is a factor of between 3.7 to 6.2
at zero nadir angle, to a factor of 2 for large nadir angles.  However, due to the 
particularly low normalization of the TD\_SLSC flux, the kilometer-size detector would 
not be sufficient for its detection.  
Fig. 15 shows the GRB\_WB model in which the enhancement factor is between
1.5 to 2, depending on energy threshold and angle.  The event rates for showers 
with energies above 10 TeV are comparable with the background, but higher 
energy threshold of 100 TeV 
would still give a few events per year for large nadir angle with 
negligible 
background.  

Fig. 16a) and 16b) show the shower event rates for the atmospheric
$\nu_\mu+\bar{\nu}_\mu+\nu_e+\bar{\nu}_e$ and $\nu_\tau+\bar{\nu}_\tau$
fluxes, respectively. For showers with 
energies above 10 TeV, 
the event rates are twice as large as for the
$E^{-1}$ flux at small nadir angle.
For $E_{\rm shr}^{\rm min}=10$ TeV,
we find the event rates for the showers to be
about 8-18 per km$^3$ per year per steradian
for the $E^{-2}$ flux,
compared with the atmospheric background of 2-16.
  
The AGN rates will stand out above the atmospheric background for 
$E_{\rm shr}^{\rm min}\sim 10$ TeV. The GRB\_WB rates are more than 
half of the atmospheric
neutrino rates at the 10 TeV shower threshold at small nadir angles. 
The TD rates are all quite low overall, and in comparison to the atmospheric
background rates.

Since one does not measure separately the tau neutrino induced
shower rates and the muon and electron induced shower rates, given 
a particular model, one can compare the rates with the
oscillation hypothesis to the predicted rates without oscillations.
We discuss the rates for specific models here, then discuss
a more model independent analysis in the next section.
To illustrate the effect of oscillations we plot in Figs. 17-20 the ratio 
of the shower event rates from $\nu_\tau$, $\nu_\mu$ and $\nu_e$  
in the oscillation scenario to the shower rates from $\nu_\mu$ 
and $\nu_e$ in the standard model, with no oscillations.  

From Fig. 17a) we note that for the  $E^{-1}$ flux, 
the shower event rates in the oscillation scenario are
a factor of 3.3-3.7 larger than in the no oscillation case 
for $E_{\rm shr}^{\rm min}=1-100$ TeV for $\theta=0^{\circ}$.
They are a factor of 1.6 enhanced for the horizontal shower rate.  
For the $E^{-2}$ 
flux, the enhancement is a factor 
of 1.4-1.6 relative to the no oscillation case for $E_{\rm shr}^{\rm min}=1-100$ TeV,
shown in Fig. 17b).
In the case of AGN models, if one assumes oscillations, the shower event 
rates are factor of 1.8-2.1 larger for AGN\_SS at zero nadir angle, decreasing
to 1.5 for nearly horizontal showers, as shown in Fig. 18a). 
Fig. 18b) shows a
ratio ranging between 1.4-1.9 for AGN\_M95 for small nadir angles.  

From Fig. 19a), we note that the shower event rates for TD\_WMB
are factor of 1.8-2.1 enhanced for energy thresholds of 1-100 TeV
for the upward neutrinos, while the 
enhancement is a factor of 1.5 for almost horizontal showers.  
In the TD\_SLSC model, Fig. 19b), the shower event rate 
is a factor of 3-3.6 enhanced at small nadir angles,
and a factor of 1.6 enhanced for horizontal showers. 
In the case of GRB\_WB model, Fig. 20 shows an
enhancement of
1.4-1.7 for 
$E_{\rm shr}^{\rm min}=1-100$
TeV. 

\subsection{Relative Rates}

We have shown that comparison 
of the muon and shower rates serves as a diagnostic for
$\nu_\mu\leftrightarrow \nu_\tau$ oscillations over astronomical
distances. For example, for the $E^{-2}$ flux,
\begin{equation}
{\rm Ratio}\Bigl[(\nu_\tau+\nu_\mu+\nu_e\rightarrow{\rm shower})_{\rm osc}
/(\nu_\mu+\nu_e\rightarrow{\rm shower})_{\rm no-osc}\Bigr]\simeq 1.5
\end{equation}
while
\begin{equation}
{\rm Ratio}\Bigl[(\nu_\tau+\nu_\mu\rightarrow\mu)_{\rm osc}
/(\nu_\mu\rightarrow\mu)_{\rm no-osc}\Bigr]\simeq 0.5\ .
\end{equation}
The ratios include contributions to showers and muons from antineutrinos.  
This feature of a deficit of muon rates and an excess of shower rates
in the oscillation scenario compared to the no-oscillation scenario
is a generic feature of all the neutrino spectra in Fig. 1.
To demonstrate this point quantitatively, we define a ratio of ratios,
\begin{equation}
{R}\equiv{
{\rm Ratio}\Bigl[(\nu_\tau+\nu_\mu+\nu_e\rightarrow{\rm shower})_{\rm osc}
/(\nu_\mu+\nu_e\rightarrow{\rm shower})_{\rm no-osc}\Bigr]\over
{\rm Ratio}\Bigl[(\nu_\tau+\nu_\mu\rightarrow\mu)_{\rm osc}
/(\nu_\mu\rightarrow\mu)_{\rm no-osc}\Bigr]
}\ .
\end{equation}

As Figs. 6-9 and 17-20 illustrate for individual fluxes, $R$
depends on energy threshold and angle. 
We show in Fig. 21 the band of $R$ spanned by the
representative models of Fig. 1 for three thresholds in muon or
shower energy: a) 1 TeV, b) 10 TeV and c) 100 TeV. 
We note that $R$ depends on nadir angle and threshold energy, 
however,
${ R}
\mathrel{\raisebox{-.6ex}{$\stackrel{\textstyle>}{\sim}$}} 2.4$ 
independent of the initial flux.  
For a given model, 
measured rates will be very distinct from predicted rates 
if the SuperK results for oscillation parameters are
correct.

Determining 
\begin{equation}
R_{exp}\equiv {
{\rm Ratio}\Bigl[({\rm shower\ rate})_{\rm measured}
/{\rm (\mu\ rate)}_{\rm measured}\Bigr]\over
{\rm Ratio}\Bigl[
(\nu_\mu+\nu_e\rightarrow{\rm shower})_{\rm no-osc}
/(\nu_\mu\rightarrow\mu)_{\rm no-osc}\Bigr]
}
\end{equation}
nevertheless relies on theoretical input for the ``no-oscillation'' flux.
Different energy behaviors of incident fluxes will have implications for
the angular and energy dependence of the event rates of upward muons and
upward hadronic/EM showers, allowing for an indirect characterization of the
energy dependence of the source. 

A more model independent test of the oscillation scenario
would be to compare the measured ratio of showers to muons 
with the no-oscillation predictions, on an absolute
scale. This requires
a crude separation of the different energy behaviors
of the fluxes of Fig. 1. By only looking at the ratio of shower to muon
events,  one could confuse the AGN\_M95 
no-oscillation ratio with the similar GRB\_WB
oscillation ratio. However, experimentally, the energy and angular
dependence of the muon event rates for the two fluxes are quite 
different, 
and GRB neutrinos would reveal themselves by time correlations 
to observed GRB events.  
One category of fluxes, with not too steep energy behavior
($E^{-1}$, AGN\_SS, TD\_WMB, TD\_SLSC and GRB\_WB), have a
reduction in the event rates from a muon threshold of 1 TeV to 10 TeV
at nadir angle of $0^\circ$ by a factor
of less than 4, whereas for 
the other two steeper fluxes ($E^{-2}$ and AGN\_M95),
the reduction is by more than a factor of 6. These reduction factors are
independent of whether or not oscillations occur.
The steeper fluxes are also
distinguished by muon event rates with a less marked dependence on nadir angle.
If one separates the steep from the less steep examples used here,
then at all three threshold energies, the band of oscillation 
shower to muon ratios does not 
 overlap with the band of no-oscillation ratios.
This is shown graphically in Figs. 22 (a-f) for 1 TeV, 10 TeV and 100 TeV
thresholds, respectively. In fact, for the 100 TeV threshold, 
one does not need any information about
the energy dependence of the initial flux, but in this case, 
the event rates are expected to be low.  

\section{Discussion}

We have studied signals for $\nu_\mu \rightarrow \nu_\tau$ oscillations 
with 
extragalactic high energy 
muon neutrinos.  Assuming SuperK oscillation parameters, muon neutrinos 
convert into tau 
neutrinos as they travel megaparsec distances, with both fluxes being 
equal at the surface of 
the Earth.  High energy muon neutrinos get absorbed as they pass through 
the Earth, while 
tau neutrinos cascade down to lower energies.  We find this enhancement 
of the $\nu_\tau$ flux 
in the low energy region to be prominent for flat initial spectrum, such 
as $E^{-1}$, the AGN model of Stecker and Salamon, 
and the topological model of Sigl {\it et al}.  For steeper spectra, the 
enhancement is small because 
the number of higher energy neutrinos that contributes to the lower energy 
flux via tau decay is 
relatively small compared to the low energy flux of neutrinos.
  
Upward tau neutrinos, once they reach the detector, interact producing 
tau leptons which decay 
with very short lifetimes.  We have considered muons from tau decay as 
well as its hadronic decay mode.  
Since the planned detectors are unable to distinguish between hadronic 
and electromagnetic showers, 
we have included all the processes that give both hadronic and 
electromagnetic showers.  We find that 
upward muons alone would not be sufficient to separate the tau neutrinos 
contribution, due to the 
large background from $\nu_\mu$ charged-current interactions, the small 
branching fraction for 
$\tau\rightarrow \mu$ decay mode and the model uncertainty for the
incident neutrino flux.  

In the case of upward hadronic/EM showers, we find 
that tau 
neutrinos give significant contributions, signaling the 
$\nu_\tau$ appearance. Given the uncertainties in the normalizations
of the extragalactic neutrino fluxes, combining muon rates and hadronic/EM
rates offer the best chance to test the $\nu_\mu\rightarrow \nu_\tau$ 
oscillation hypothesis.  

As concluded in earlier work \cite{gqrs96,gqrs98}, in general, 
an energy threshold of between 10 TeV and  100 TeV for 
upward muons and showers 
is needed in order to reduce the background from atmospheric neutrinos.  
We find that diffuse AGN neutrino fluxes, as described by the Stecker-Salamon 
and Mannheim models, 
as well as 
neutrinos from GRBs can be used to detect tau appearance.  
By measuring upward showers with
energy threshold of 10 TeV, and upward muons, the event rates exceed 
the atmospheric
background and are about a factor of 1.5-2 larger than in the no-oscillation 
scenario.

Here we also comment on the 
effect of muon and pion cooling to the flavor ratio.  
Athar {\it et al.} in Ref. \cite{yasuda} have shown that with a negligible
electron neutrino content at the source, the electron neutrino content
at the Earth (in the three-flavor model) is reduced if 
not negligible compared
to the nearly equal muon and tau neutrino fluxes. Keeping the energy spectrum
unchanged, this means that the hadronic/electromagnetic shower background,
which has significant contributions from $\nu_e N\rightarrow e X$ with
$\nu_e>E_{\rm shr}^{\rm min}$ would be reduced. Electron (anti-) neutrinos
from processes in the propagation of cosmic rays
may dominate at some energies \cite{stanev0003}. 
We have not considered that possibility here
because of the low rates below 1 PeV.

Steepening of the energy spectra displayed in 
Fig. 1 due to a neutrino energy cutoff from pion and muon cooling
will have implications for the
tau neutrino `pileup', especially 
for the flatter spectra where the pileup is more
pronounced. As an estimate of the lower bound on the relative
enhancement of the hadronic/EM signal compared to the muon signal,
one can compare the rates for horizontal events, where tau neutrino pileup
is small. 
For example, 
Figures 22 (a-f) show clear distinction between oscillation and
no-oscillation scenarios, even in directions near horizontal, where there is 
no pile up.  Furthermore, for 
$E^{-2}$ flux, where the pileup is very small \cite{irs}, the 
ratio of ratios $R$ discussed above ranges from 2.5 to 2.8.  
Thus, even without the tau neutrino pileup, the 
oscillation scenario can be distinguished from the 
no-oscillation scenario.  

The detection of 
$\nu_\mu \rightarrow \nu_\tau$ oscillations with a point source might also 
be possible.  With the 
resolution for the planned neutrino telescopes of $2^{\circ}$, the atmospheric
background is reduced by $3.8 \times 10^{-3}$.  
For upward showers, this gives less than 1 event per year for 
$E_{\rm shr}^{\rm min}=1$ TeV, and even less 
for higher energy thresholds.  Thus, 
if the point source has a flat spectrum, 
$F_{\nu+\bar{\nu}} = 10^{-16} E^{-1}$, 
then one would be able to detect tau neutrinos by measuring upward 
showers with $E_{\rm shr}^{\rm min}=1$ TeV.  
In the more realistic case, when the point source has a steeper spectrum
($E^{-2}$), 
such as Sgr A* \cite{markoff98}, a normalization of $10^{-7}$/cm$^2$/s/sr/GeV 
would be sufficient for the
detection of tau neutrinos with threshold of 1 TeV. Time correlations with
variable point sources would further enhance 
the signal relative to the background.

We have demonstrated that extragalactic sources of neutrinos can be used 
as a very-long baseline experiment, providing a source of tau neutrinos
and opening up a new frontier in studying neutrinos oscillations.   

\vskip 0.1true in
\leftline{Acknowledgements}  

The work of S.I.D. and I.S. has been supported in part
by the DOE under Contracts 
DE-FG02-95ER40906 and DE-FG03-93ER40792.  The work of M.H.R. has been 
supported in 
part by
National Science Foundation Grant No.
PHY-9802403.

\appendix\section{Tau decay distribution}

The decay distribution of the tau neutrinos from 
tau decay has the following form,
in terms of $z = E_\nu/E_\tau$:
\begin{equation}
{dn\over dz}=\sum_i B_i(g_0^i+P\,g_1^i)\ .
\end{equation}
The polarization of the decaying $\tau^-$ is $P$, which for neutrino
V-A production of $\tau^-$ is $P=-1$. The branching fraction into
decay channel $i$ is indicated by $B_i$. The distribution is normalized
such that
\begin{equation}
\int {dn\over dz}dz =\sum_i B_i = 1\ .
\end{equation}
In Table 1, we show the functions $g_0$ and $g_1$ for each decay mode,
written in terms of $z$ and $r_i=m_i^2/m_\tau^2$.
Details of the calculational procedure can be found in Ref. \cite{gaisserbook}
or in Ref. \cite{liparimu}.
For the multiprong tau decays, we approximate the distribution by
a theta function, as indicated in the table.

\newpage
\begin{table}
\caption{Functions $g_0$ and $g_1$ in the tau neutrino 
energy ($E_\nu$) distribution
from $\tau$ decays, in terms of $z= E_\nu/E_\tau$. Note,
$X$ indicates hadrons, with  $X\neq \pi,\ \rho, \ 
a_1.$}
\begin{tabular}{lccc}
Process & $B_\tau$ & $g_0$ & $g_1$ \\ 
& & & \\ \hline
& & & \\
$\tau\rightarrow \nu_\tau \mu \nu_\mu$ & 0.18  &  ${5\over 3} - 
3z^2+{4\over 3}z^3$
& ${1\over 3} - 3z^2+{8\over 3}z^3 $\\
$\tau\rightarrow \nu_\tau e \nu_e $ & & & \\
& & &\\
$\tau\rightarrow \nu_\tau \pi$ & 0.12 & ${1\over 1-r_\pi}\theta(1-r_\pi-z)$ 
& $-{2z -1+r_\pi\over (1-r_\pi)^2}\theta(1-r_\pi-z)$ \\
& & &\\
$\tau\rightarrow \nu_\tau \rho$ & 0.26 & ${1\over 1-r_\rho}\theta(1-r_\rho-z)$ 
& $-\biggl({2z-1+r_\rho\over 1-r_\rho}\biggr)\biggl({1-2r_\rho\over 1+2r_\rho}
\biggr)\theta(1-r_\rho-z)$\\
& & &\\
$\tau \rightarrow  \nu_\tau a_1$ & 0.13 & ${1\over 1-r_{a_1}}
\theta(1-r_{a_1}-z)$ 
& $-\biggl({2z-1+r_{a_1}\over 1-r_{a_1}}\biggr)
\biggl({1-2r_{a_1}\over 1+2r_{a_1}}
\biggr)\theta(1-r_{a_1}-z)$\\
& & & \\
$\tau\rightarrow \nu_\tau X$ & 0.13 & ${1\over 0.3}\theta(0.3-z)$ & 0 \\
& & & \\
\end{tabular}
\end{table}
\newpage

\begin{figure}[!hbt]
\rule{0.0cm}{1.0cm}\vspace{1.0cm}\\
\epsfxsize=17cm
\epsfbox[0 0 4096 4096]{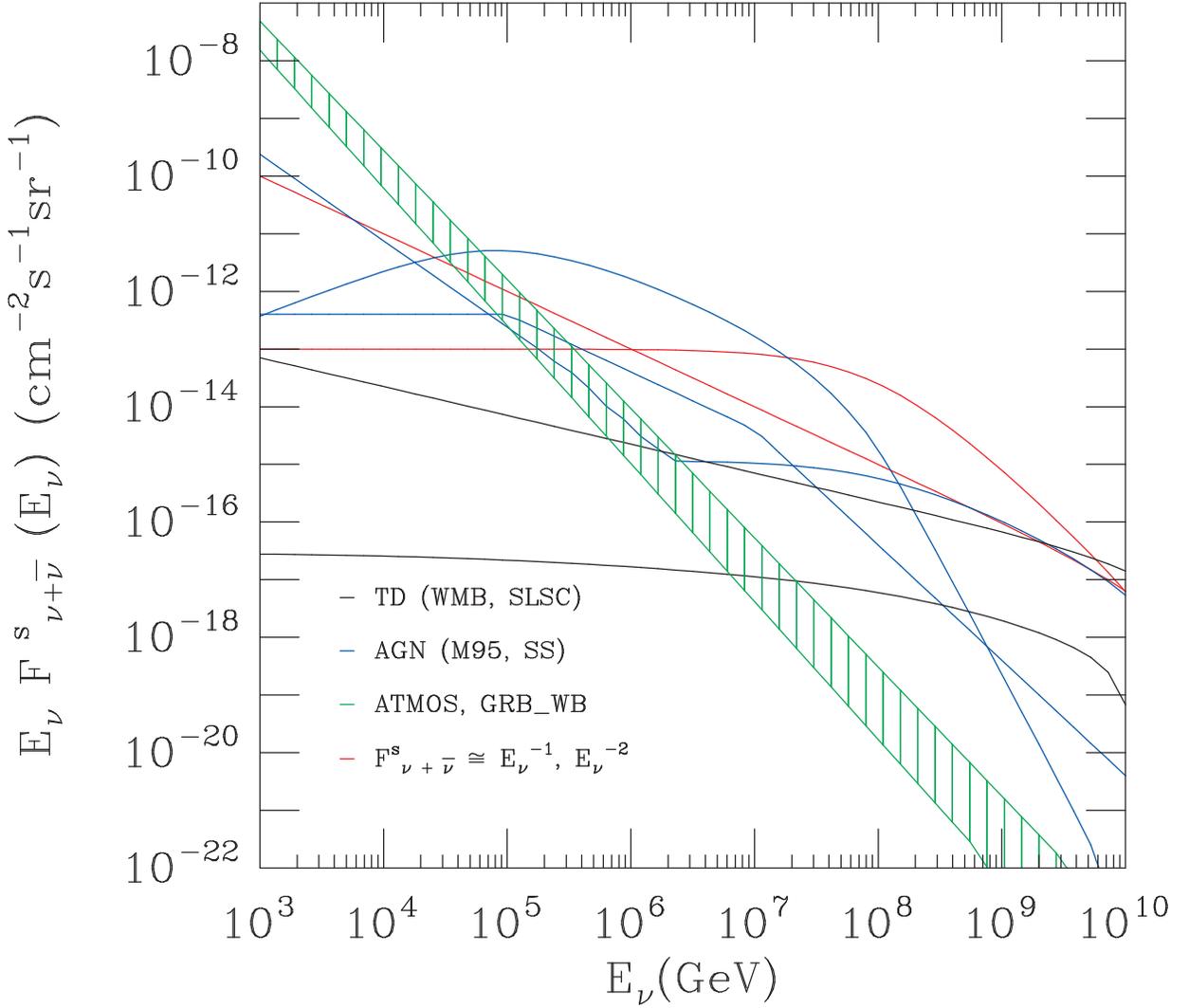}
\medskip
\rule{0.0cm}{0.1cm}\vspace{-2.0cm}\\
\caption{\normalsize{Muon neutrino plus antineutrino fluxes for AGN models (blue 
lines, upper curve 
at low energy corresponds to AGN\_M95, while the 
lower curve is for AGN\_SS model), 
GRB (green line), topological defects models (black lines, upper curve corresponds to 
the TD\_WMB, while
the lower curve is for TD\_SLSC), 
$E^{-1}$ flux (lower red line at low energy) and $E^{-2}$ (upper red line
at low energy)
and angle-dependent atmospheric (ATM)
flux (green shaded area). }}
\end{figure}
\newpage
\newpage
\begin{figure}[!hbt]
\rule{0.0cm}{1.0cm}\vspace{2.0cm}\\
\epsfxsize=15cm
\epsfbox[0 0 4096 4096]{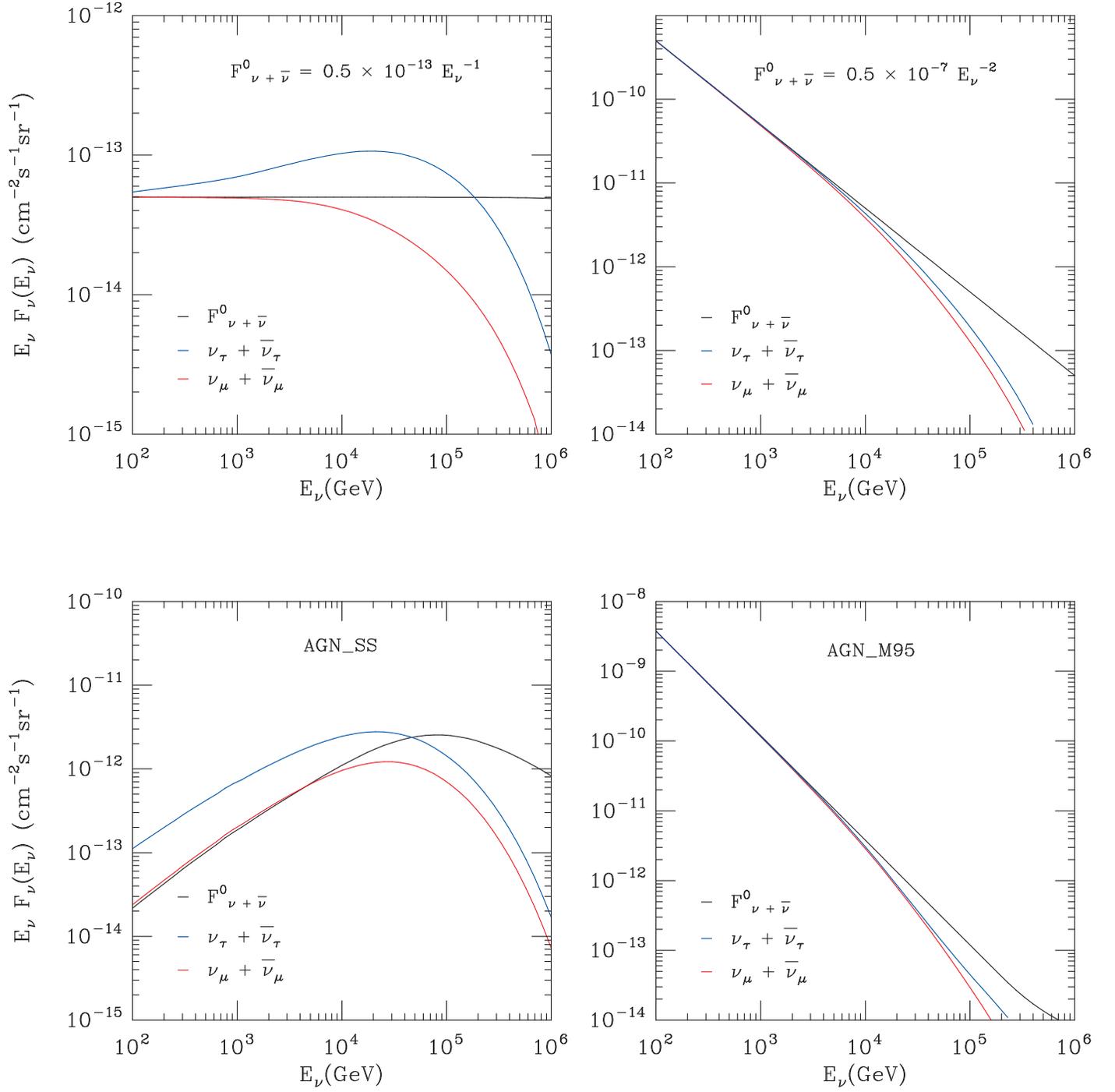}
\medskip
\rule{0.0cm}{0.1cm}\vspace{0.1cm}\\
\caption{\normalsize{Muon neutrino plus antineutrino flux (black line), 
the effect of its attenuation 
for $\theta=0^{\circ}$ (red line) and tau neutrino plus antineutrino 
upward flux for the same initial flux and the same nadir angle 
(blue line) for
a) $E^{-1}$ flux b) $E^{-2}$ flux c) AGN\_SS and d) AGN\_M95.}}
\end{figure}
\newpage

\begin{figure}[!hbt]
\rule{0.0cm}{1.0cm}\vspace{2.0cm}\\
\epsfxsize=15cm
\epsfbox[0 0 4096 4096]{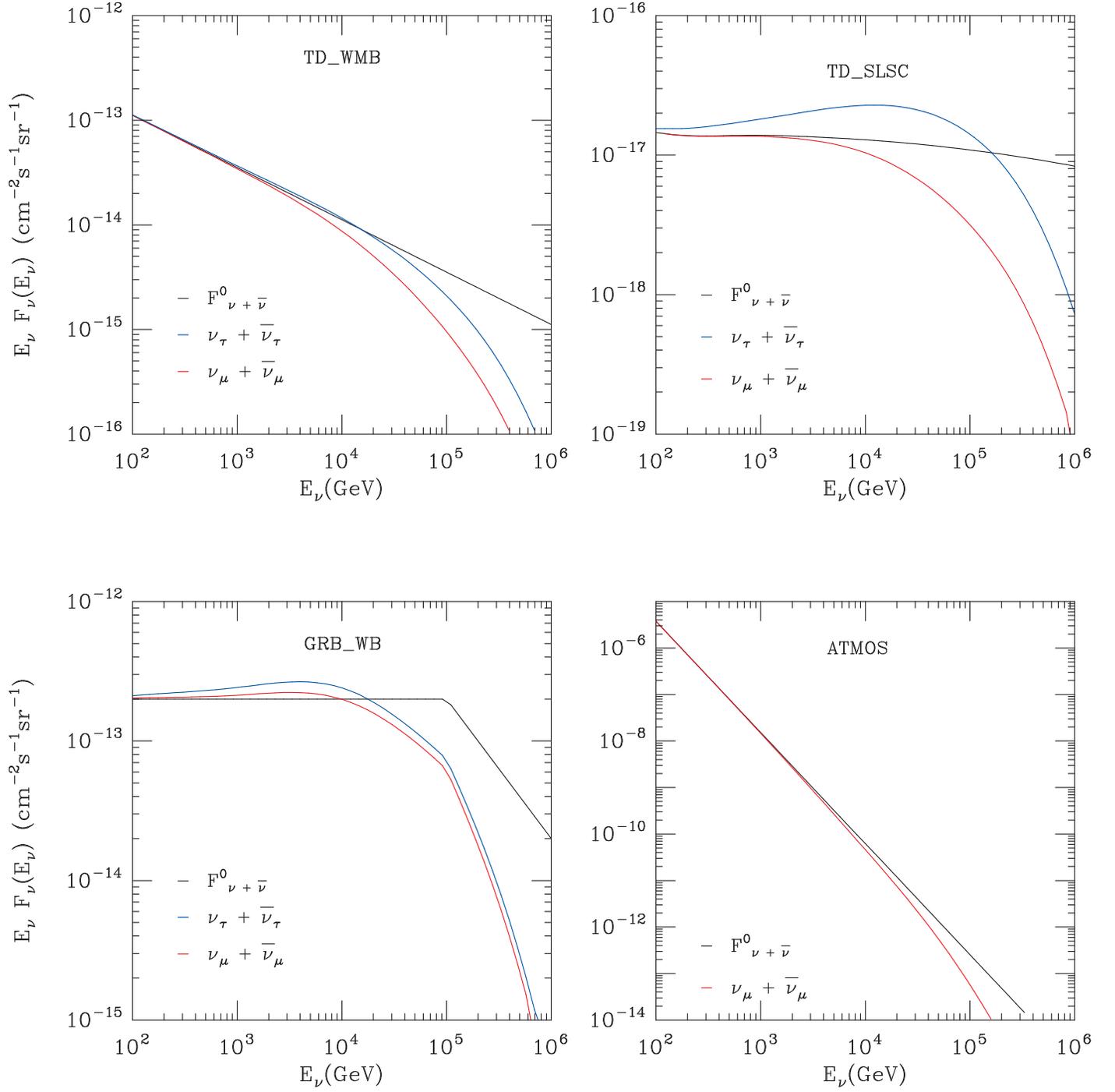}
\medskip
\rule{0.0cm}{0.1cm}\vspace{0.1cm}\\
\caption{\normalsize{Muon neutrino plus antineutrino flux (black line), 
the effect of its attenuation 
for $\theta=0^{\circ}$ (red line) and tau neutrino plus antineutrino 
upward flux for the same initial flux and the same nadir angle 
(blue line) for
a) TD\_WMB, b) TD\_SLSC, c) GRB\_WB and d) the atmospheric flux ATMOS.}}
\end{figure}
\newpage

\begin{figure}[!hbt]
\rule{0.0cm}{1.0cm}\\
\epsfxsize=15.5cm
\epsfbox[0 0 4096 4096]{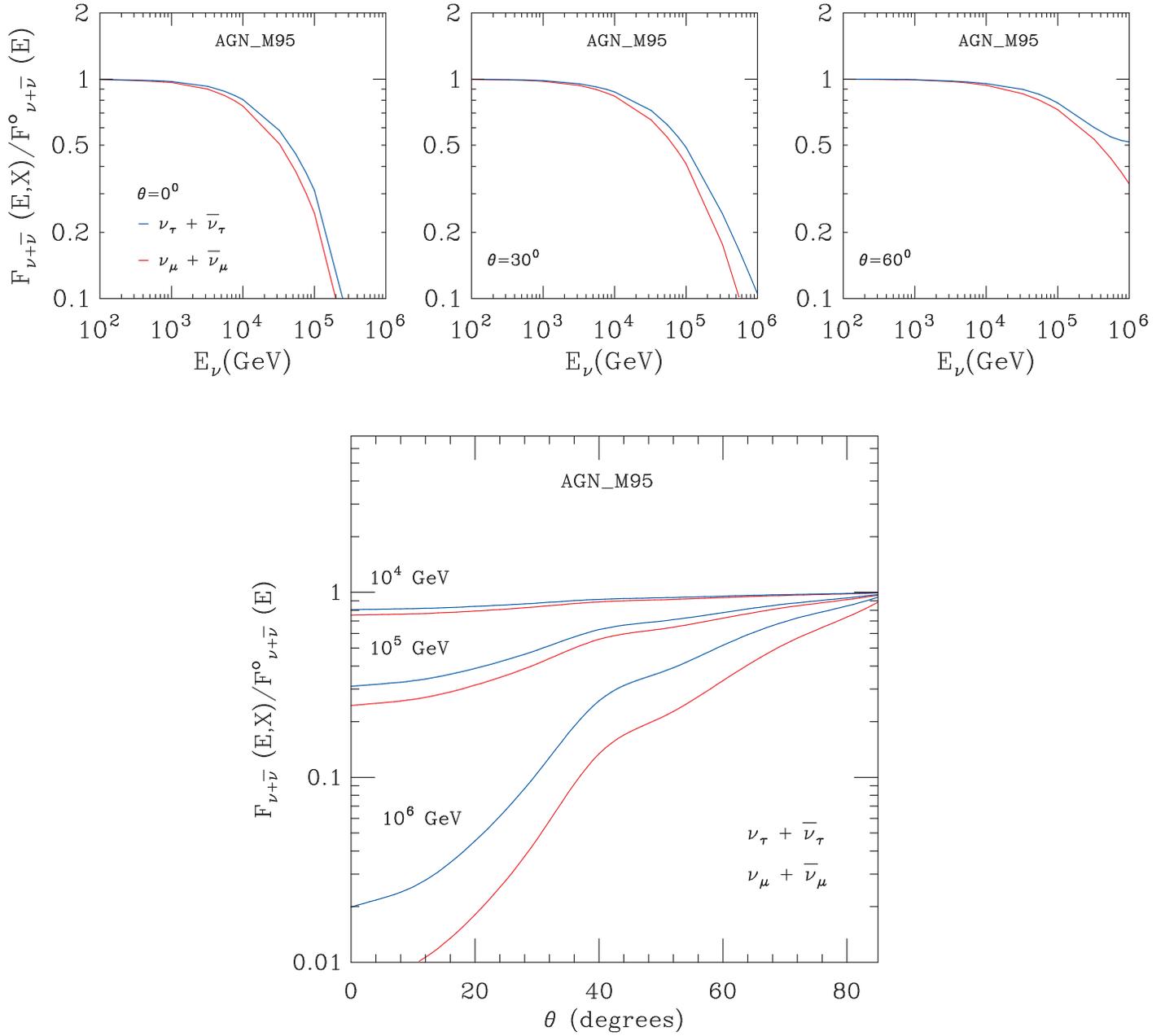}
\medskip
\rule{0.0cm}{0.1cm}\vspace{3.0cm}\\
\caption{ The energy dependence of the muon neutrino flux (red line) and the 
tau neutrino flux (blue line) for nadir angle $\theta=0^{\circ}$, 
$\theta=30^{\circ}$ and $\theta=60^{\circ}$ normalized to the initial flux 
for the AGN\_M95.}
\end{figure}
\newpage

\begin{figure}[!hbt]
\rule{0.0cm}{1.0cm}\vspace{-12.0cm}\\
\epsfxsize=20cm
\epsfbox[0 0 700 700]{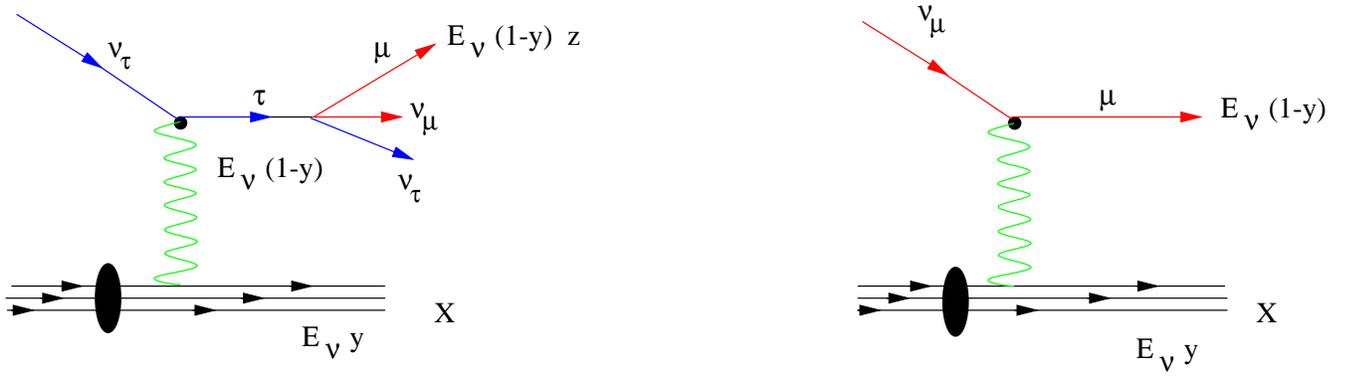}
\medskip
\rule{0.0cm}{0.1cm}\vspace{1.0cm}\\
\caption{\normalsize{Diagrams for neutrino interactions contributing to 
the muon production.}}
\end{figure}

\newpage
\begin{figure}[!hbt]
\rule{0.0cm}{1.0cm}\vspace{3.0cm}\\
\epsfxsize=12.cm
\epsfbox[0 0 4096 4096]{Menrat12.eps}
\medskip
\rule{0.0cm}{0.1cm}\vspace{3.cm}\\
\caption{\normalsize{Muon event rate as a function of nadir angle for energies 
1 TeV, 10 TeV and 100 TeV. Muon rates including the 
contribution from tau decay (blue line) compared
with the background from muon neutrinos (red line) for 
a) $E^{-1}$ flux and b) $E^{-2}$ flux.}}
\end{figure}
\newpage
\begin{figure}[!hbt]
\rule{0.0cm}{1.0cm}\vspace{2.0cm}\\
\epsfxsize=12cm
\epsfbox[0 0 4096 4096]{MenratSSMN.eps}
\medskip
\rule{0.0cm}{0.1cm}\vspace{3.0cm}\\
\caption{\normalsize{Muon event rate as a function of nadir angle for energies 
1 TeV, 10 TeV and 100 TeV. Muon rates including the 
contribution from tau decay (blue line) compared
with the background from muon neutrinos (red line) for 
a) AGN\_SS and b) AGN\_M95.
}}
\end{figure}
\newpage

\begin{figure}[!hbt]
\rule{0.0cm}{1.0cm}\vspace{2.0cm}\\
\epsfxsize=12cm
\epsfbox[0 0 4096 4096]{MenratTDLC.eps}
\medskip
\rule{0.0cm}{0.1cm}\vspace{3.0cm}\\
\caption{\normalsize{Muon event rate as a function of nadir angle for energies 
1 TeV, 10 TeV and 100 TeV. Muon rates including 
contribution from tau decay (blue line) compared
with the background from muon neutrinos 
(red line) for a) TD\_WMB and b) TD\_SLSC.
}}
\end{figure}
\newpage

\begin{figure}[!hbt]
\rule{0.0cm}{1.0cm}\vspace{2.0cm}\\
\epsfxsize=12cm
\epsfbox[0 0 4096 4096]{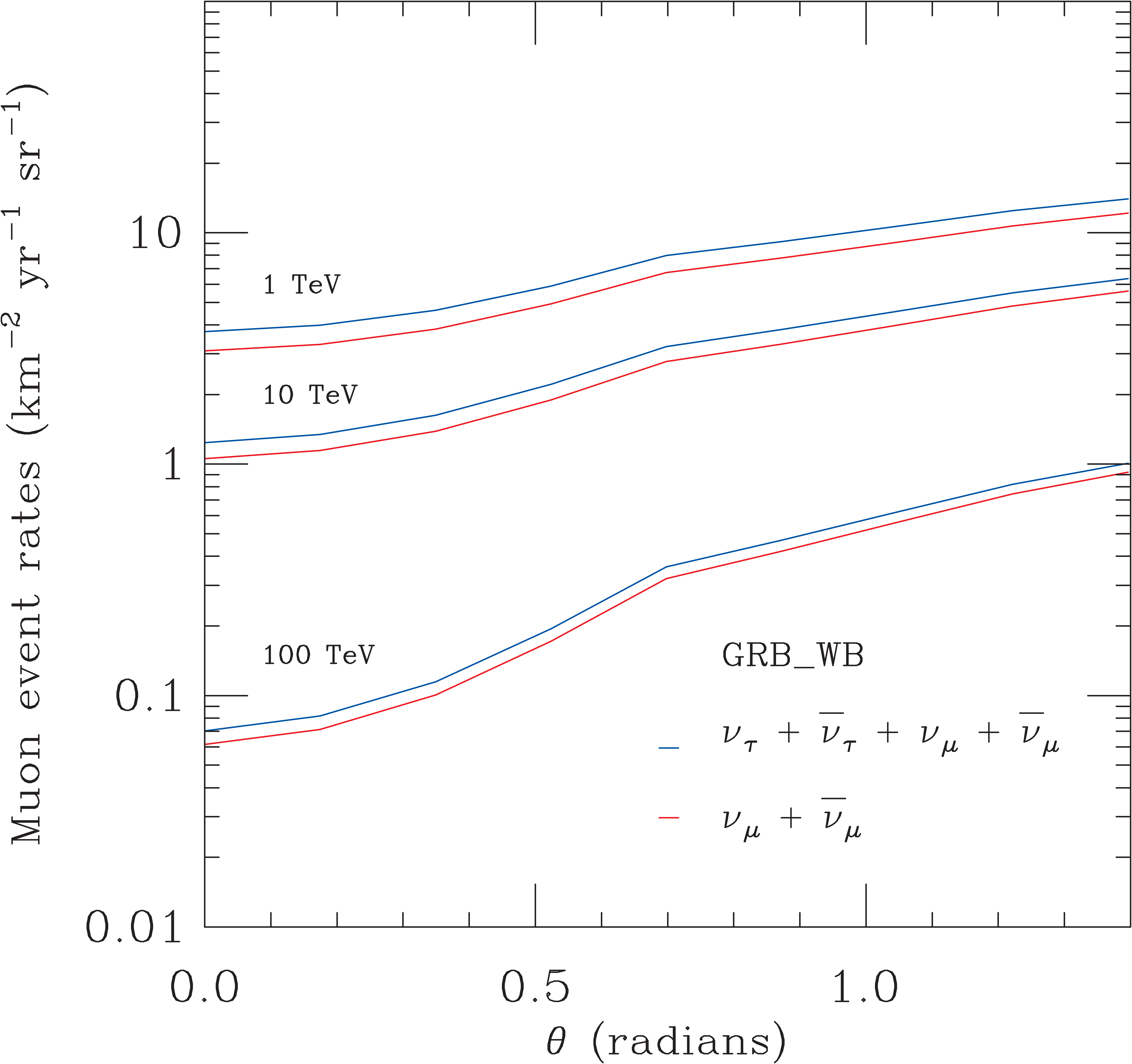}
\medskip
\rule{0.0cm}{0.1cm}\vspace{4.0cm}\
\caption{\normalsize{Muon event rate as a function of nadir angle for energies 
1 TeV, 10 TeV and 100 TeV. Muon rates including the 
contribution from tau decay (blue line) compared
with the background from muon neutrinos 
(red line) for GRB\_WB.}}
\end{figure}
\newpage

\begin{figure}[!hbt]
\rule{0.0cm}{1.0cm}\vspace{2.0cm}\\
\epsfxsize=12cm
\epsfbox[0 0 4096 4096]{MenratAT.eps}
\medskip
\rule{0.0cm}{0.1cm}\vspace{4.0cm}\
\caption{\normalsize{Muon event rate as a function of nadir angle for energies 
1 TeV, 10 TeV and 100 TeV for  a) Atmospheric muon neutrino and b) 
Atmospheric tau neutrino.}}
\end{figure}
\newpage

\begin{figure}[!hbt]
\rule{0.0cm}{1.0cm}\vspace{-9.0cm}\\
\epsfxsize=20cm
\epsfbox[0 0 1000 1000]{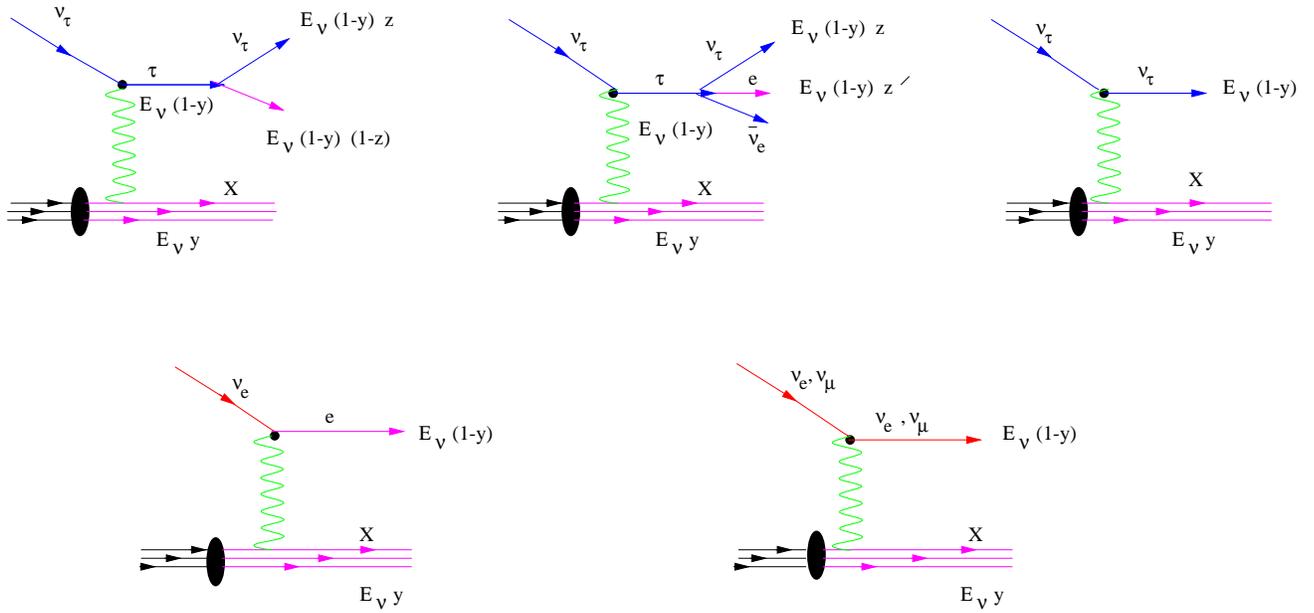}
\medskip
\rule{0.0cm}{0.1cm}\vspace{3.0cm}\\
\caption{\normalsize{Diagrams for neutrino interactions contributing 
to the shower events.}}
\end{figure}

\newpage
\begin{figure}[!hbt]
\rule{0.0cm}{1.0cm}\vspace{2.0cm}\\
\epsfxsize=12cm
\epsfbox[0 0 4096 4096]{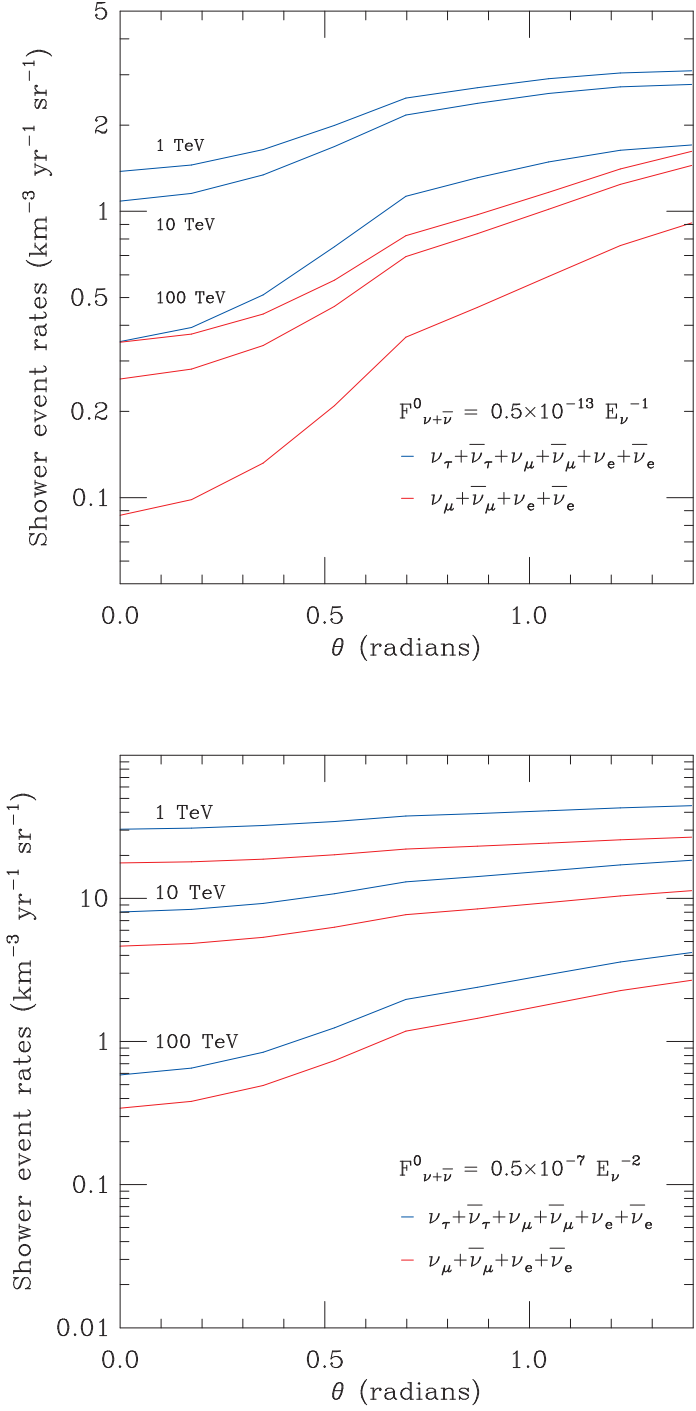}
\medskip
\rule{0.0cm}{0.1cm}\vspace{4.0cm}\\
\caption{\normalsize{Hadronic/EM event rates as a 
function of nadir angle for $E_{\rm shr}^{\rm min}=$1 TeV, 
10 TeV and 100 TeV. 
Hadronic/EM event rates from $\nu_\tau$ (blue line) compared
hadronic/EM event rates from $\nu_\mu$ plus $\nu_e$ (red line) for 
a) $E^{-1}$ and b) $E^{-2}$. }}
\end{figure}
\newpage 

\begin{figure}[!hbt]
\rule{0.0cm}{1.0cm}\vspace{1.0cm}\\
\epsfxsize=12cm
\epsfbox[0 0 4096 4096]{EenSSMN.eps}
\medskip
\rule{0.0cm}{0.1cm}\vspace{4.0cm}\\\\
\caption{\normalsize{Hadronic/EM event rates as a function of nadir angle 
for $E_{\rm shr}^{\rm min}=$1 TeV, 10 TeV and 100 TeV.  
Hadronic/EM event rate from
$\nu_\tau$ (blue line) compared
hadronic/EM event rate from $\nu_\mu$ plus $\nu_e$ 
(red line) for a) AGN\_SS and b) AGN\_M95.
}}
\medskip
\end{figure}
\newpage

\begin{figure}[!hbt]
\rule{0.0cm}{1.0cm}\vspace{2.0cm}\\
\epsfxsize=12cm
\epsfbox[0 0 4096 4096]{EenTDLC.eps}
\medskip
\rule{0.0cm}{0.1cm}\vspace{4.0cm}\\
\caption{\normalsize{Hadronic/EM event rates as a function of nadir angle for 
$E_{\rm shr}^{\rm min}=$ 
1 TeV, 10 TeV and 100 TeV.  
Hadronic/EM event rates from $\nu_\tau$ (blue line) 
compared
hadronic/EM event rates from $\nu_\mu$ plus $\nu_e$ 
(red line) for a) TD\_WMB and b) TD\_SLSC.
}}
\medskip
\end{figure}

\newpage
\begin{figure}[!hbt]
\rule{0.0cm}{1.0cm}\vspace{2.0cm}\\
\epsfxsize=12cm
\epsfbox[0 0 4096 4096]{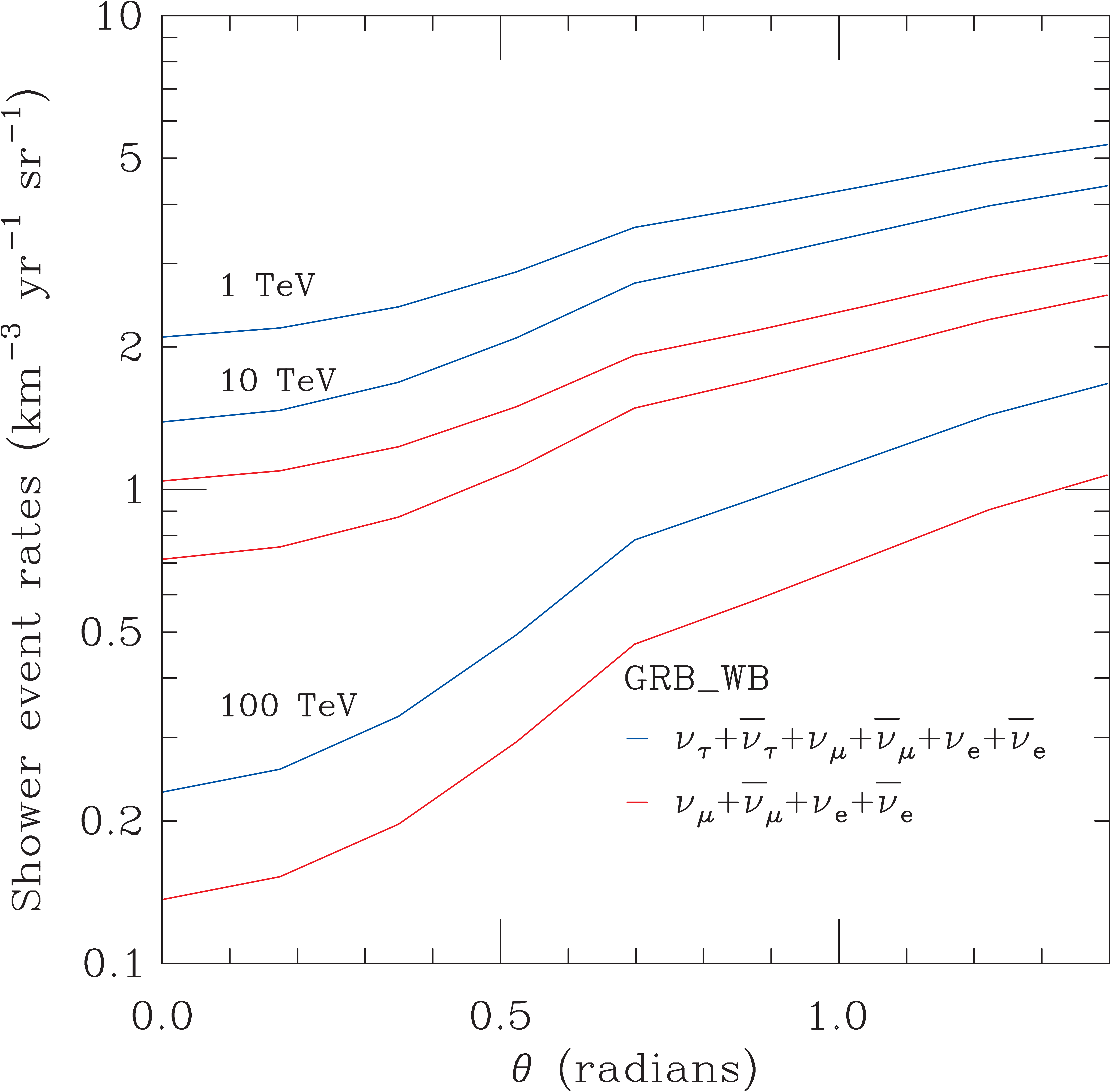}
\medskip
\rule{0.0cm}{0.1cm}\vspace{4.0cm}\\
\caption{\normalsize{Hadronic/EM event rates as a function of nadir angle for 
$E_{\rm shr}^{\rm min}=$ 
1 TeV, 10 TeV and 100 TeV. 
Hadronic/EM event rates from $\nu_\tau$ (blue line) compared
hadronic/EM event rates from $\nu_\mu$ plus $\nu_e$ 
(red line) for GRB\_WB. 
}}
\medskip
\end{figure}
\newpage
\begin{figure}[!hbt]
\rule{0.0cm}{1.0cm}\vspace{2.0cm}\\
\epsfxsize=12cm
\epsfbox[0 0 4096 4096]{EenAT.eps}
\medskip
\rule{0.0cm}{0.1cm}\vspace{4.0cm}\\
\caption{\normalsize{Hadronic/EM event rates as a function of nadir angle for 
$E_{\rm shr}^{\rm min}=$ 
1 TeV, 10 TeV and 100 TeV. 
Hadronic/EM event rates from $\nu_\tau$ (blue line) compared
hadronic/EM event rates from $\nu_\mu$ plus $\nu_e$ 
(red line) for a) atmospheric muon neutrino and b) atmospheric tau 
neutrino.}}
\medskip
\end{figure}

\newpage
\rule{0.0cm}{0.1cm}\vspace{1.0cm}\\
\begin{figure}[!hbt]
\rule{0.0cm}{1.0cm}\\
\epsfxsize=12cm
\epsfbox[0 0 4096 4096]{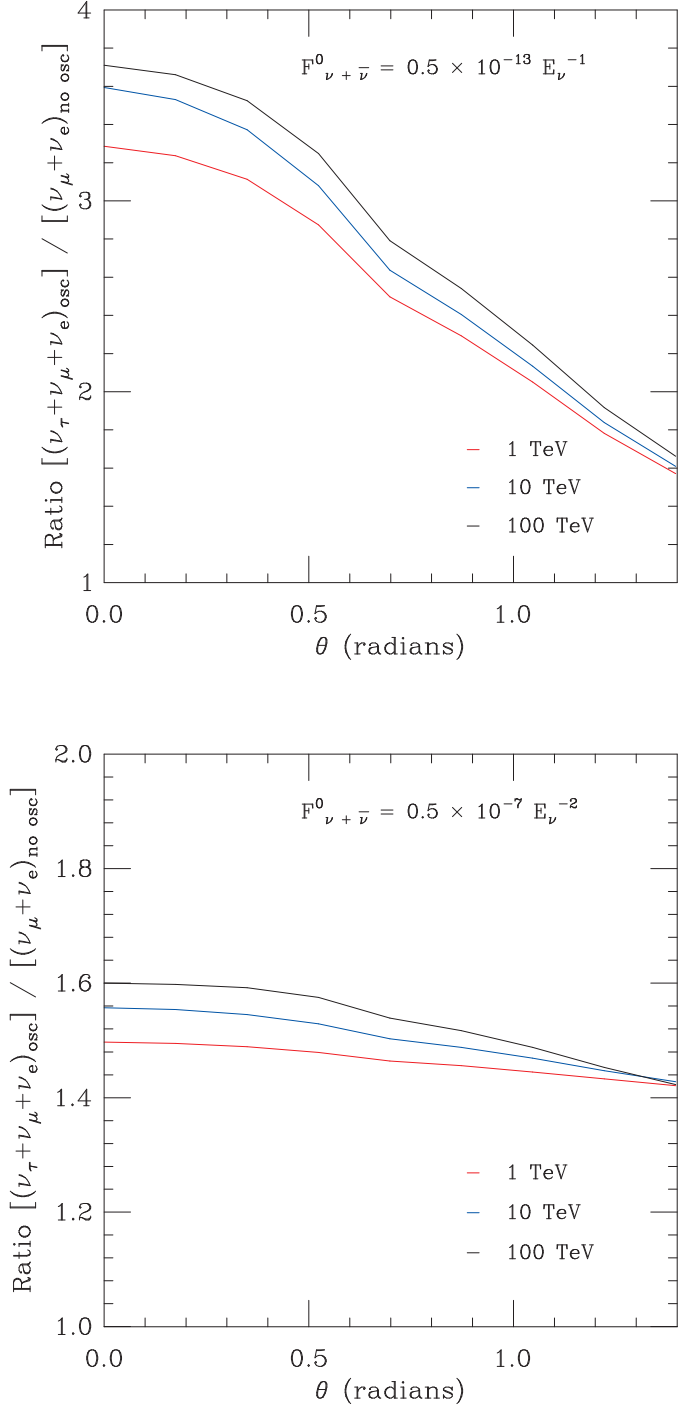}
\medskip
\rule{0.0cm}{0.1cm}\vspace{2.8cm}\\
\caption{\normalsize{Ratio of Hadronic/EM event rate of 
$\nu_\tau$ plus $\nu_\mu$ plus $\nu_e$ assuming oscillation scenario and 
$\nu_\mu$ plus $\nu_e$ in the standard model 
as a function of nadir angle for energies 1 TeV, 10 TeV and 100 TeV for 
a) $E^{-1}$  and 
b) $E^{-2}$ flux.}}
\end{figure}
\newpage
\begin{figure}[!hbt]
\rule{0.0cm}{1.0cm}\vspace{2.0cm}\\
\epsfxsize=12cm
\epsfbox[0 0 4096 4096]{RenSSMN.eps}
\medskip
\rule{0.0cm}{0.1cm}\vspace{4.0cm}\\
\caption{\normalsize{Ratio of Hadronic/EM event rates of 
$\nu_\tau$ plus $\nu_\mu$ plus $\nu_e$ assuming oscillation scenario and 
$\nu_\mu$ plus $\nu_e$ in the standard model 
as a function of nadir angle for energies 1 TeV, 
10 TeV and 100 TeV for a) AGN\_SS and 
b) AGN\_M95 flux.}}
\end{figure}
\newpage
\vspace{2in}
\begin{figure}[!hbt]
\rule{0.0cm}{1.0cm}\vspace{2.0cm}\\
\epsfxsize=12cm
\epsfbox[0 0 4096 4096]{RenTDLC.eps}
\medskip
\rule{0.0cm}{0.1cm}\vspace{4.0cm}\\
\caption{\normalsize{Ratio of Hadronic/EM event rate of 
$\nu_\tau$ plus $\nu_\mu$ plus $\nu_e$ assuming oscillation scenario and 
$\nu_\mu$ plus $\nu_e$ in the standard model 
as a function of nadir angle for $E_{\rm shr}^{\rm min}=$1 TeV, 
10 TeV and 100 TeV for 
a) TD\_WMB and 
b) TD\_SLSC flux.}}
\end{figure}
\newpage
\begin{figure}[!hbt]
\rule{0.0cm}{1.0cm}\vspace{3.0cm}\\
\epsfxsize=12cm
\epsfbox[0 0 4096 4096]{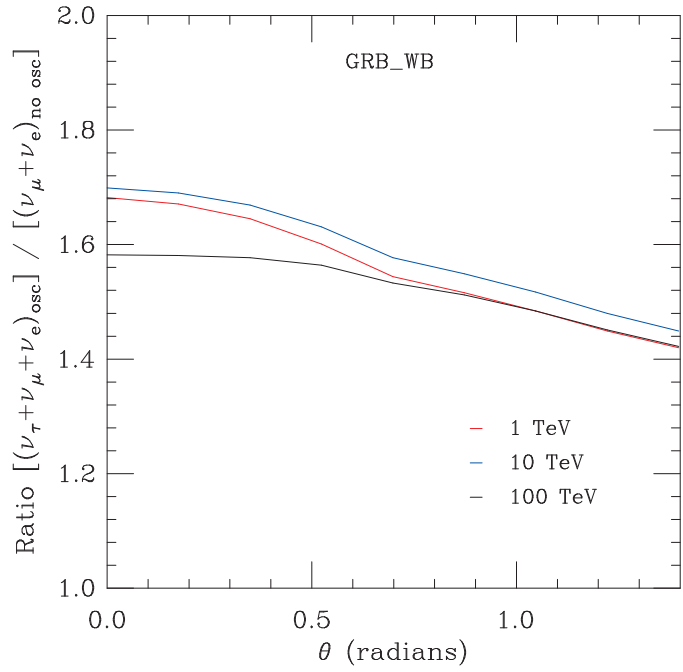}
\medskip
\rule{0.0cm}{0.1cm}\vspace{4.0cm}\\
\caption{\normalsize{Ratio of Hadronic/EM event rate of 
$\nu_\tau$ plus $\nu_\mu$ plus $\nu_e$ assuming oscillation scenario and 
$\nu_\mu$ plus $\nu_e$ in the standard model 
as a function of nadir angle for energies 1 TeV, 
10 TeV and 100 TeV for GRB\_WB.}}
\end{figure}
\begin{figure}[!hbt]
\rule{0.0cm}{1.0cm}\vspace{3.0cm}\\
\epsfxsize=12cm
\epsfbox[0 0 3696 3696]{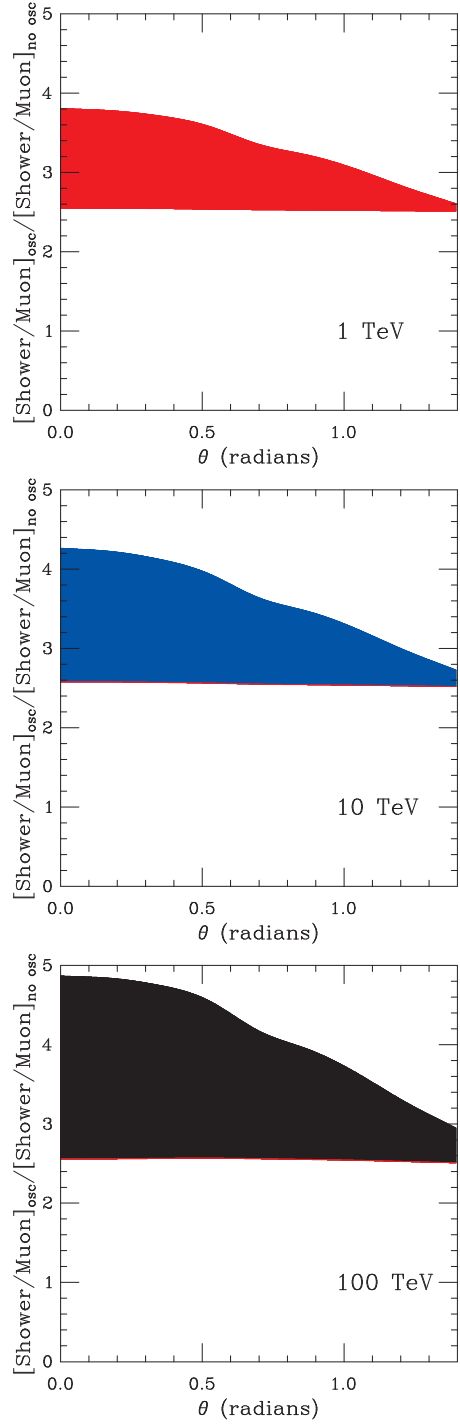}
\medskip
\rule{0.0cm}{0.1cm}\vspace{4.0cm}\\
\caption{\normalsize{Ratio of Hadronic/EM event rate of 
$\nu_\tau$ plus $\nu_\mu$ plus $\nu_e$ assuming oscillation scenario and 
$\nu_\mu$ plus $\nu_e$ in the standard model relative to the ratio
of the muon event rate of $\nu_\tau$ plus $\nu_\mu$ assuming oscillation
scenario and $\nu_\mu$ in the standard model (see Eq. (15))
as a function of nadir angle for threshold energies 1 TeV, 10 TeV 
and 100 TeV.
}}
\end{figure}
\begin{figure}[!hbt]
\rule{0.0cm}{1.0cm}\vspace{3.0cm}\\
\epsfxsize=12cm
\epsfbox[0 0 3696 3696]{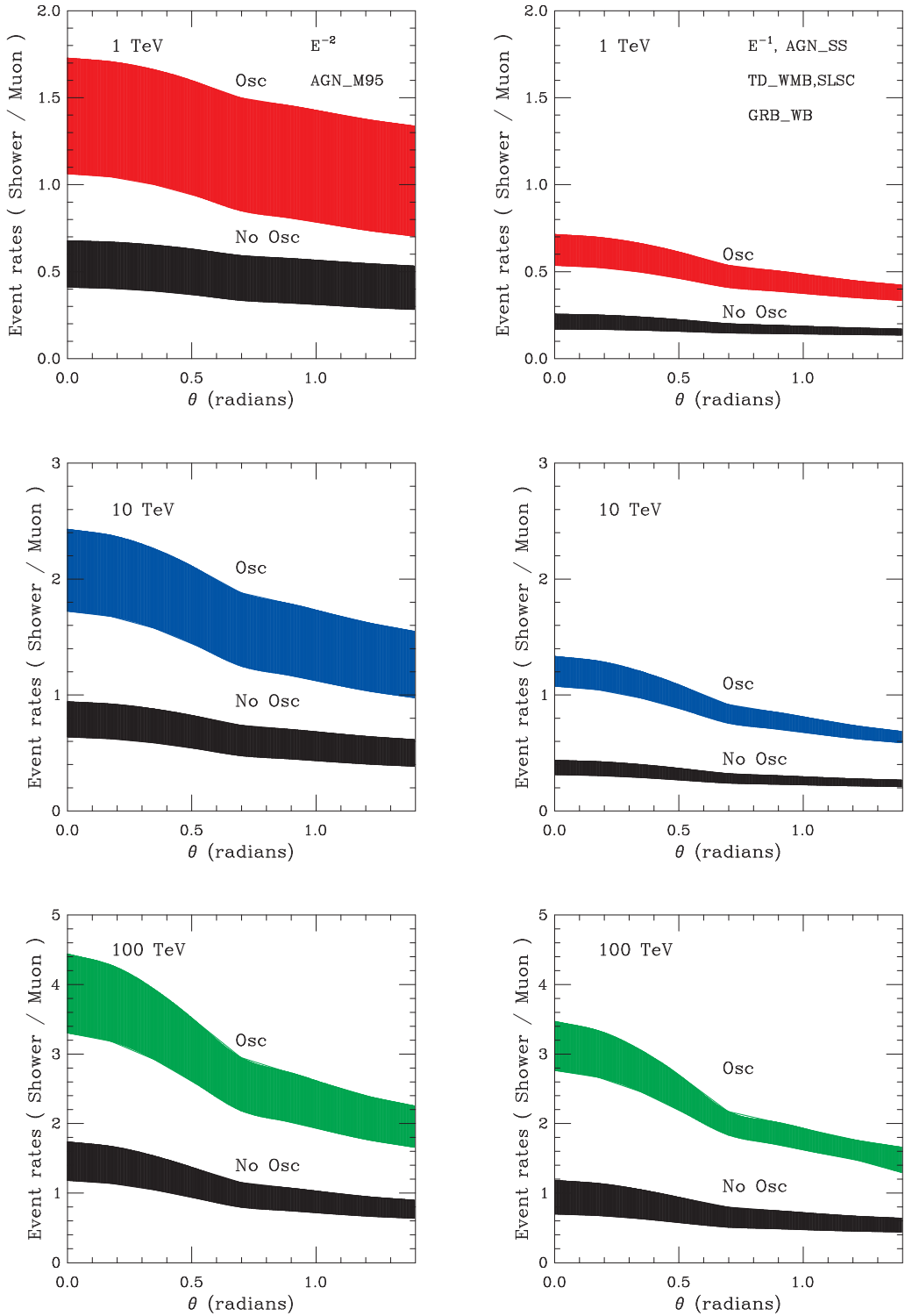}
\medskip
\rule{0.0cm}{0.1cm}\vspace{4.0cm}\\
\caption{\normalsize{Ratio of Hadronic/EM event rate to muon event
rate for the oscillation and no-oscillation scenarios
as a function of nadir angle for threshold energies (a-b) 1 TeV, 
(c-d) 10 TeV and 
(e-f) 100 TeV for the indicated fluxes.
}}
\end{figure}

\end{document}